\documentclass[useAMS,usenatbib, use]{mn2e}
\usepackage{epsfig, color, amsmath, bm, subfig}
\title{Bayesian Asteroseismology of 23 Solar-Like Kepler Targets}
\author[Gruberbauer et al.]{M. Gruberbauer$^{1}$\thanks{E-mail: mgruberbauer@ap.smu.ca}, 
D.B. Guenther$^{1}$\thanks{E-mail: guenther@ap.smu.ca},
K. MacLeod$^{1}$\thanks{E-mail: kmacleod@mars.smu.ca}
T. Kallinger$^{2}$\thanks{E-mail: thomas.kallinger@univie.ac.at}
\\
$^{1}$Institute for Computational Astrophysics, Department of Astronomy and Physics, Saint Mary's University, Halifax, NS B3H 3C3, Canada\\
$^{2}$Institute for Astrophysics, University of Vienna, T\"urkenschanzstrasse 17, A-1180 Vienna, Austria \\}

\newcommand{\XO}{X_{\rm 0}}
\newcommand{\ZO}{Z_{\rm 0}}
\newcommand{\YO}{Y_{\rm 0}}

\newcommand{\Zs}{Z_{\rm s}}

\newcommand{\ZoX}{Z_s/X_s}

\newcommand{\aml}{\alpha_{\rm ml}}
\newcommand{\muHz}{\mu{\rm Hz}}

\begin{document}

\date{Accepted  Received}

\pagerange{\pageref{firstpage}--\pageref{lastpage}} \pubyear{2013}

\maketitle

\label{firstpage}

\begin{abstract}
We study 23 previously published Kepler targets to perform a consistent grid-based Bayesian asteroseismic analysis and compare our results to those obtained via the Asteroseismic Modelling Portal (AMP). We find differences in the derived stellar parameters of many targets and their uncertainties. While some of these differences can be attributed to systematic effects between stellar evolutionary models, we show that the different methodologies deliver incompatible uncertainties for some parameters. Using non-adiabatic models and our capability to measure surface effects, we also investigate the dependency of these surface effects on the stellar parameters. Our results suggest a dependence of the magnitude of the surface effect on the mixing length parameter which also, but only minimally, affects the determination of stellar parameters. While some stars in our sample show no surface effect at all, the most significant surface effects are found for stars that are close to the Sun's position in the HR diagram.
\end{abstract}


\section{Introduction}
\label{sec:intro}
Ultra-precise long-term photometric time series from space have revolutionized the study of stellar variability in recent years. The CoRoT \citep{michel2008} and the {\it Kepler} \citep{borucki2010} space telescopes in particular have produced high-quality data sets for thousands of stars in order to detect planets down to Earth size and below. Particularly interesting for the study of stellar interiors and stellar evolution is their ability to detect solar-type oscillations from giants to subdwarfs. The pulsational characteristics of these stars adhere, at least to a very good first approximation, to scaling relations \citep[e.g.,][]{huber2011} permitting the study of large populations of stars with ``ensemble asteroseismology" \citep{Chaplin2011} and even Galactic archeology \citep[][]{Miglio2013}.

The same information can also be exploited to infer the parameters of individual stars, e.g., to better constrain their planets' properties. For stellar astrophysics, however, the ultimate goal is to use asteroseismology to study stellar interiors. Instead of direct inversion of the pulsation information, asteroseismology usually employs a comparison between observed and modelled pulsation frequencies \citep[e.g.,][]{guenther2004}. Various new tools have been developed to facilitate a state-of-the-art version of such a comparison using different approaches, such as the Asteroseismic Modelling Portal (AMP) \citep{metcalfe2009} and Bayesian grid-based analysis \citep[][hereafter Paper I]{gruberbauer2012}. The major differences between these methods lie in their different statistical basis and their different applications of what is known as the surface effect correction (see Paper I for an in-depth discussion). Already, the AMP has been used to analyse some Kepler targets in detail and to compare the results with those from other modellers \citep{metcalfe2010, metcalfe2012}. Such a comparison is advantageous, because asteroseismic modelling often relies on a specific set of stellar models with a specific set of input physics. Slight systematic differences among these models are therefore not only plausible but unavoidable, resulting in underestimated uncertainties. A different approach is to study a larger sample of stars self-consistently with one particular method and model base to facilitate a pool of results to be compared with other researcher's results \citep{mathur2012}.

In this paper we reexamine some of the previously published studies based on Kepler data with a strong emphasis on AMP results, employing our own set of models and our Bayesian method described in Paper I. We will discuss how the results differ and whether the methodologies themselves introduce systematic deviations. We will also perform the first detailed study on surface effects for a sample of stars with our flexible method. 

\section{Methods, models and observations}

\subsection{Target selection and observations}
In order to investigate the impact of the stellar models and methodologies in the most general sense, we analyse stars for which the p-mode frequency sets and detailed prior information used in previous asteroseismic fitting procedures are available in the literature. We furthermore constrain ourselves to stars that do not show strong signatures of deviations from the asymptotic relation, i.e., avoided crossings such as in KIC 11026764 \citep{metcalfe2010}. While those signatures are very valuable for asteroseismic inferences and can be easily taken into account with our method as mentioned in Paper I, they would constitute special cases in the comparison between methods. We therefore postpone such an analysis to a future paper and restricted ourselves to 20 of the 22 stars analysed by \cite{mathur2012} (hereafter Mathur20), the solar analogues 16 Cyg A \& B \citep{metcalfe2012}, and the planet-host Kepler-36 \citep{carter2012}. Where available in the previously cited papers, we use prior constraints on $\log T_{\rm eff}$, $\log L/L_{\odot}$, $Z/X$ (adopting $[{\rm Fe/H}]_{\odot} = 0.0245$) and $\log g$. Following our description in Paper I, these prior constraints are modelled as separate Gaussian probability distributions. 

As is common in recent asteroseismic analyses, we treat the frequency of maximum power $\nu_{\rm max}$ as an additional and independent observable by using the scaling relation

\begin{equation}
\nu_{\rm max, mod} \approx \frac{M/M_{\odot}\left(T_{\rm eff}/T_{\rm eff, \odot}\right)^{3.5}}{L/L_{\odot}}\nu_{\rm max, \odot}.
\label{equ:nu_max}
\end{equation} 
where we employ the solar value $\nu_{\rm max, \odot} = 3120.0 \pm 5 \,\muHz$ given by \cite{kallinger2010c} based on VIRGO data. For Mathur20, the observed values and uncertainties of $\nu_{\rm max}$ have been taken from \cite{mathur2012}, and for Kepler-36 we have used the value published in \cite{carter2012}. For 16 Cyg A \& B, we have determined the values ourselves by performing a Bayesian multi-component model fit, consisting of a flat background, three super-Lorentzian profiles and a Gaussian power hump, to the power density spectra of both data sets\footnote{Note that only Q7 data obtained between September 2010 to December 2010 have been used in \cite{metcalfe2012}. We therefore restrict ourselves to this data set as well.}. In this case, the central frequency of the Gaussian power hump and the corresponding uncertainties are interpreted as a good proxy for $\nu_{\rm max}$. The method employs the nested sampling algorithm MultiNest \citep{feroz2009} and is described in more detail in \cite{kallinger2010c}. We find $\nu_{\rm max} = 2215.6 \pm 6.5\,\muHz$ for 16 Cyg A, and $\nu_{\rm max} = 2571.9 \pm 12.6\,\muHz$ for 16 Cyg B.

\subsection{Models}

A wide parameter range has to be spanned in order to perform a meaningful grid-based analysis. We therefore employed YREC \citep{demarque2008} to produce a set of dense grids covering a wide range in initial masses, and several values for the initial helium mass fraction $\YO$, initial metal mass fraction $\ZO$, and mixing length parameter $\aml$. 

Our model tracks begin as completely convective Lane-Emden spheres \citep{lane1869, chandra1957} with the stellar age reset to zero when the star crosses the birthline \citep[$10^{-5}\,\rm M_{\odot} / yr$,][]{palla1999}. They are evolved from the Hayashi track \citep{hayashi1961} through the zero-age-main-sequence (ZAMS) to the base of the red giant branch. Constitutive physics include the OPAL98 \citep{iglesias1996} and \cite{alexander1994} opacity tables, as well as the Lawrence Livermore 2005 equation of state tables \citep{rogers1986, rogers1996}. Convective energy transport was modelled using the B\"ohm-Vitense mixing-length theory \citep[MLT,]{boehm1958}. The atmosphere is implemented using Eddington gray atmosphere. Nuclear reaction cross-sections were taken from \cite{bahcall2001} and the nuclear reaction rates from Table 21 in \cite{bahcall1988}. The effects of helium and heavy element diffusion \citep{bahcall1995} were included. The model grid contains models with $M/M_{\odot}$ from 0.8 to 1.3 in steps of 0.01 and $\YO$ from 0.210 to 0.315 in steps of 0.005. Furthermore, $\ZO$ is varied from 0.005 to 0.04 in steps of 0.005, but with the overall constraint that $\XO \geq 0.68$. Lastly, we also vary $\aml$ from 1.8 to 2.4 in steps of 0.1. 

The pulsation spectra were computed using the stellar pulsation code of \cite{guenther1994}, which solves the linearized, non-radial, non-adiabatic pulsation equations using the Henyey relaxation method. The non-adiabatic solutions include radiative energy gains and losses but do not include the effects of convection. We estimate the random $1\sigma$ uncertainties of our model frequencies to be of the order of $0.1\,\mu\rm Hz$. These uncertainties are properly propagated into all further calculations.

\subsection{Bayesian asteroseismic grid fitting}
\label{paper3:sec:fitting}

Our Bayesian fitting method is explained in detail in Paper I, and it has been previously applied to analyse the Sun \citep{gruberbauer2013}. We compare theoretical ($f_{\rm m}$) and observed ($f_{\rm o}$) frequencies by calculating the likelihood that the two values agree were it not for the presence of random and systematic errors, i.e.,
\begin{equation}
f_{\rm o} - f_{\rm m} = \gamma \Delta + e.
\end{equation}
Here, the random errors $e$ are assumed to be independent and Gaussian. The systematic errors $\gamma \Delta$ in the case of solar-like stars are assumed to be similar to ``surface effects". At higher orders, observed frequencies are systematically lower than model frequencies, and the absolute frequency differences increase with frequency. This is modelled by introducing $\Delta$ as free parameters for each observed mode and by setting $\gamma = -1$. 

These $\Delta$ terms are then allowed to become larger at higher radial orders. The upper limit $\Delta_{\rm max}$ for each model frequency $f_{\rm m}$ is determined by the large frequency separation and a power law similar to the standard correction introduced by \cite{kjeldsen2008} so that
\begin{equation}
\Delta_{\rm max} = \Delta\nu \left(\frac{f_{\rm m}}{f_{\rm max, m}}\right)^b,
\label{equ:dNuscaling}
\end{equation} 
where $b = 4.9$, $\Delta\nu$ is the large frequency separation of the corresponding model, and $f_{\rm max, m}$ is the frequency of the highest order in the model\footnote{This means that for the highest order in the model $\Delta_{\rm max} = \Delta\nu$ and guarantees that we do not introduce ambiguities in the radial orders by implementing the $\Delta$ terms.}. 

The $\Delta$ parameter is incorporated in a completely Bayesian fashion, using a $\beta$ prior to prefer smaller values over larger ones (see Paper I for more details). In addition, we always allow for the possibility that a mode is not significantly affected by any kind of systematic error by explicitly including the null hypothesis, that is by combining the probabilities of two hypotheses: one with and one without the $\Delta$ parameter. Altogether, this allows us to fully propagate uncertainties originating from the surface effects, or other potential systematic differences, into all our results. At the same time it gives us more flexibility than the standard surface-effect correction. Whereas the latter prescribes a fixed power-law behaviour for the actual surface effects, our method only prescribes such a behaviour for the {\it upper limits} of the surface effects for the individual radial model frequencies.

For each model in our grid, all the likelihood terms from the different frequencies are combined to yield an overall weighted likelihood for the model, where the weights are provided either via prior information or using ignorance priors (i.e., information that simply encodes the dimension of the grid). These weights provide correctly normalized probabilities that allow us to derive distributions for all model properties (e.g., mass, age, fractional radius of the base of the convection zone $R_{\rm BCZ}$, mixing-length parameter $\alpha_{\rm ML}$, and so on).

In summary, we obtain probabilities for every evolutionary track in our grid, and within the tracks also for every model. We also obtain the correctly propagated distributions for systematic errors so that the model-dependent surface effect can be measured. In order to fully resolve the changes in stellar parameters and details in the stellar-model mode spectra, we oversample the evolutionary tracks via linear interpolation until the (normalised) probabilities no longer change significantly. Eventually, we obtain so-called evidence values, equivalent to the prior-weighted average likelihood, for the grid as a whole. These could, in principle, be used to perform a quantitative evaluation of different input physics \citep[see][]{gruberbauer2013} or even different stellar evolution and pulsation codes. We will use them in this study to analyse the significance of the measured surface effects.

In order to facilitate this, we propose two alternative systematic error models in addition to the standard surface effect (SSE) model described above. First, we employ a less restrictive systematic error model where 
\begin{equation}
\Delta_{\rm max} = \Delta \nu /2
\end{equation}
for every frequency of each particular model. Furthermore, the observed frequencies are allowed to deviate in either direction (first $\gamma = 1$ is evaluated, then $\gamma = -1$ follows, and then both results are combined using the sum rule). We call this model the ``arbitrary systematic error" (ASE) model since it allows, in principle, very large differences between observed and calculated frequencies without prescribing any systematic behaviour or preferred sign. Note that this is not equivalent to simply increasing the Gaussian uncertainties of the observed frequencies to $\Delta \nu /2$. 

Finally, we will also employ a third error model which only consists of the probabilities obtained without any $\Delta$ parameters. This model therefore assumes that no systematic errors are present so that $f_{\rm o} - f_{\rm m} = e$. 
We will call this the ``no systematic error" (NSE) model. Together, the three systematic error models will allow us to estimate the significance of surface effects or other systematic differences between observed and calculated frequencies. 

\section{Dependence of surface effects on non-adiabaticity and mixing length}
\label{sec:mixinglength}

The advantage of our method to include systematic frequency errors over the standard surface correction is its universality. The standard surface effect exponent of $b = 4.9$ as obtained by \cite{kjeldsen2008} has been derived for adiabatic pulsation frequencies and for a solar-calibrated model with a calibrated parametrization of the mixing length\footnote{As explained in the previous section, technically we also use $b=4.9$ for our surface effect modelling, but the exponent is only employed to derive an upper limit for the surface effect for each mode. This does not enforce the usual power-law like behaviour of the surface effect.}. More advanced pulsation models and different stellar models \citep[see, e.g.,][]{grigahcene2012} are not necessarily consistent with such a relation. This is also the case for our non-adiabatic pulsation frequencies. Since the only way to improve our modelling of outer layers is to compare more advanced models to observations, it is necessary to relax the constraint of a definite empirical surface correction relation dependent on adiabatic pulsation codes. Aside from the surface effect, our method also allows various other kind of parameterisation for systematic errors, such as our ASE model.

The drawback of our method, as discussed in Paper I, is that in the absence of strong prior information about the stellar parameters, a lack of lower-order modes will potentially result in an underestimated magnitude of the surface effects. This follows from the fact that we always obtain the most probable result given our state of information including the new data set; if we cannot constrain the stellar model parameters using our prior knowledge, the pulsation frequencies are our only reference. When conditions are encountered under which the empirical correction law of \cite{kjeldsen2008} does not apply, or if one rejects such a correction on other grounds, we have to evaluate the models acknowledging the presence of less well-specified systematic errors.

As described in Paper I, neglecting lower order modes leads to overestimated $\aml$, mass, and metallicity for the Sun, simply because such models can fit the higher order modes better. The same models cannot fit the lower order modes as well, but when they are not included in the list of fitted modes no penalty ensues. For stars other than the Sun, we usually do not have a complete list of lower-order modes, nor do we have as accurate non-seismic constraints (e.g., mass, luminosity, age).   
Even in such cases, however, stellar metallicity, $T_{\rm eff}$ and $L$ can be estimated from spectroscopic and photometric observations. Furthermore, equation\,\ref{equ:nu_max} reveals that $\nu_{\rm max}$ provides valuable if approximate constraints for the fundamental parameters, including the stellar mass, in particular when spectroscopic constraints are available. 

Two adjustable parameters of the stellar model, the helium abundance and $\aml$, affect the structure of the surface layers. The mixing length parameter is normally tuned to produce a model of the Sun at the observed composition and (meteoritic) age that matches the limb-darkening-corrected radius of the Sun. The helium abundance is either derived also from a tuned model of the Sun, matching its luminosity, or extrapolated from the observed rate of Galactic nucleosynthesis. Both the helium abundance and the $\aml$ affect the depth of the convection zone (i.e., the fitted adiabat) and the temperature gradient in the superadiabatic layer (SAL)\footnote{Below the SAL, the temperature gradient is adiabatic.} via the mixing length theory. We stress that the mixing length parameter of the MLT is used primarily to control the efficiency of convection and its adjustment is primarily used to fix the radius of the star. As is well known for the case of the Sun the MLT does not correctly predict the temperature gradients in the SAL so even though it may be providing an accurate radius for the star it may, at the same time, be providing a poor model of the SAL \citep[e.g.,][]{robinson2003}.
The surface effect is sensitive to $\aml$ since the p-mode frequencies are sensitive to the SAL. But at the same time the large frequency spacings are also sensitive to the $\aml$ via its effect on the star's radius. The interplay of the two effects of the mixing length parameter on the frequencies makes it difficult to isolate the surface effect completely from $\aml$. 
\begin{figure}
\centering
\includegraphics[width=\columnwidth]{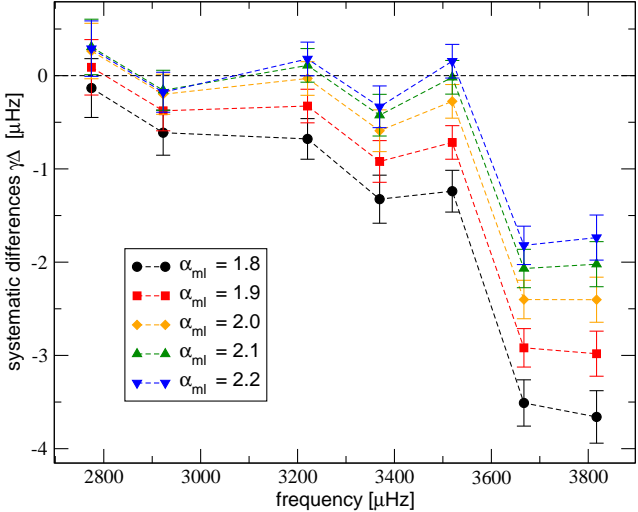}
\caption{Systematic differences between observed and computed $l=0$ modes for KIC 8006161 when fitted by models with varying mixing length but otherwise fixed initial parameters.}
\label{fig:alpha1}
\end{figure}

\begin{figure}
\centering
\includegraphics[width=\columnwidth]{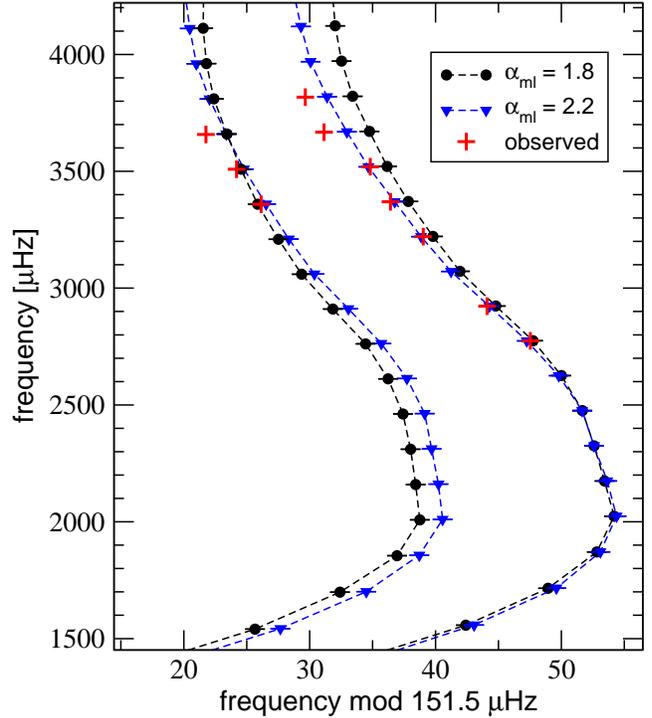}
\caption{Echelle diagrammes for the l=0 (right sequence) and 2 (left sequence) modes of KIC 8006161 and two models with different $\aml$. The uncertainties of the observed frequencies are of the order of the symbol size.}
\label{fig:alpha2}
\end{figure}

Figure\,\ref{fig:alpha1} shows the effect of fitting one of the stars in our sample, KIC 8006161, to a specific evolutionary track with $M=1.11\, M_{\odot}$, $\YO = 0.22$ and $\ZO = 0.04$, but varying values of $\aml$ (note that these models are not equivalent to our most probable models as determined in the next section). At the highest frequencies, the larger $\aml$ values clearly reduce the measured surface effect by almost 50\%, and the effect is even more pronounced at the lower orders.\footnote{It is necessary to point out, however, that for adiabatic frequencies, the relative impact of the mixing length is not as big as for the non-adiabatic frequencies.} On the other hand, the plot suggests that at the lowest observed radial order, the frequencies for the higher-$\aml$ models are somewhat too low. Figure\,\ref{fig:alpha2} presents the echelle diagramme for the l=0,2 modes of the $\aml = 1.8$ and $\aml = 2.2$ models. We observe that if the set of l=2 modes extended below $\sim 3000\,\muHz$, we would be able to clearly distinguish between these two models. At the l=0 orders below $\sim 2000\,\muHz$ the two models show small but systematic differences as well. 
With the current set of observed modes, however, we cannot clearly determine whether a lower or higher mixing length parameter value is more probable. Yet, we want to find the model with the smallest surface effects that still fits all other constraints. Therefore, in our example the higher $\aml$ values become more probable automatically. As long as we have limited knowledge on the magnitude of the surface effects across the HR diagram, however, this increase in probability might not be warranted. In the given example, it does seem as if the $\aml = 2.2$ model is more consistent with the observed small spacing, but we know that the solar-calibrated value is closer to $\aml \sim 1.8$, so deviations from this value should not be taken lightly\footnote{Note that such deviations are also a non-negligible problem when applying the standard surface correction since it relies on the solar-calibrated values at the solar mixing length parameter.}. Nonetheless, studying the possible variation of the mixing length parameter across the HRD and its interplay with the surface effects is important, so setting a fixed (calibrated) value is also not desirable. 

We therefore propose the following solution: we perform our analysis using three different approaches to constraining the mixing length. The first approach is to not use any prior on $\aml$. The second approach is to employ a Gaussian prior with $\aml = 1.8 \pm 0.075$, based on the solar calibrated value. The standard deviation of the prior (0.075) is somewhat arbitrary, but we choose it to permit deviations from the calibrated value in the presence of strong evidence. As the maximum value of $\aml$ in our grid is 2.4, such a model would represent an {\it a priori} $8\sigma$ outlier. For such a model to still be more probable, it would require differences in likelihood of about 14 orders of magnitude, and therefore a large amount of evidence from the frequencies and the fit to the other stellar parameters. The prior should therefore only lead to $\aml > 2.1$ for stars that can be matched very well both in terms of their frequencies and in terms of their fundamental parameters. Lastly, the third approach is to constrain ourselves to $\aml = 1.8$ in reference to the Sun-calibrated value for Eddington atmospheres. This set of different constraints on $\aml$ will allow us to show its impact on the stellar parameters and the surface effects. By comparing the Bayesian evidence for the result obtained with different priors, we can also quantify the formal preference of one prior over the others.
 
 \begin{figure}
\centering
\includegraphics[width=\columnwidth]{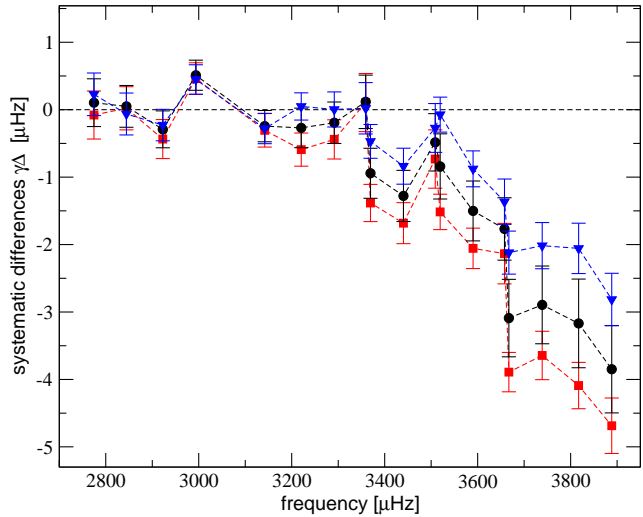}
\caption{Systematic differences between all observed and computed modes of KIC 8006161 for the whole grid, calculated with (black circles) and without (blue triangles) $\aml$ prior.
Results for only the $\aml = 1.8$ models (red squares) are also shown.}
\label{fig:alpha3}
\end{figure}

As an example, we present the results for the surface effect analysis of KIC 8006161, based on the complete grid rather than just one evolutionary track, in Figure\,\ref{fig:alpha3}. While for this star the prior does not have a big effect at the lower order modes, we obtain significantly larger surface effects beyond $3300\,\muHz$ with the $\aml$ priors. Even with the Gaussian $\aml$ prior, as will be shown below, the most probable posterior value for $\aml$ lies above 1.8. 

\section{Results}

As described in the previous sections, we have analysed all 23 stars in our sample with the same grid, using priors on their fundamental parameters if available and three different models relating to the treatment of systematic errors. Moreover, we perform this analysis three times, first setting $\aml = 1.8$, then with a Gaussian prior, and lastly without a prior on $\aml$. The results are given in Table\,\ref{tab:all1}, Table\,\ref{tab:all2}, and Table\,\ref{tab:all3}, and the most probable $\aml$ priors and surface effect models are also indicated.

\subsection{The influence of the $\aml$ priors}

\begin{figure*}
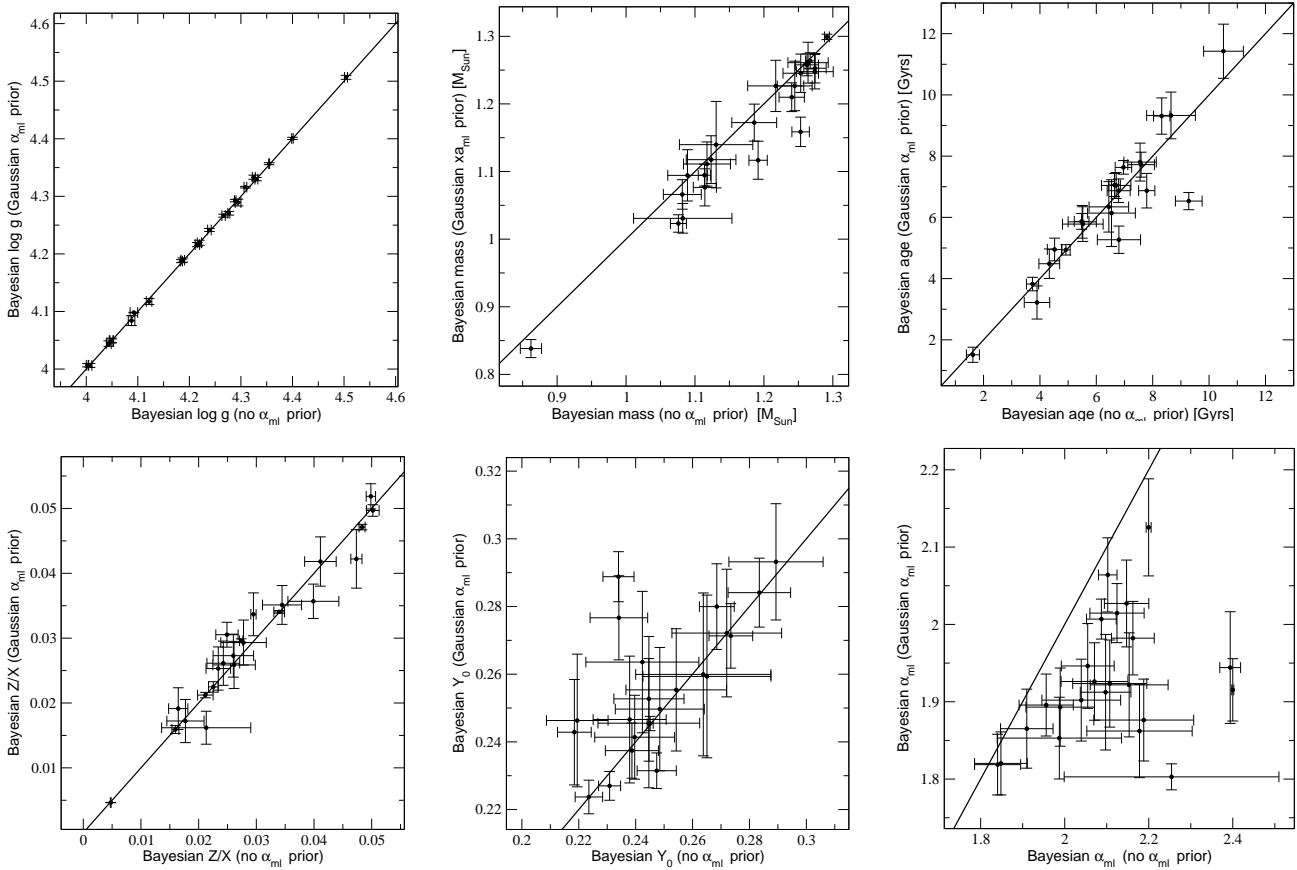

\centering
\begin{flushleft}
\subfloat{\includegraphics[width=0.3\textwidth]{figures/figure4a}}\qquad
\subfloat{\includegraphics[width=0.3\textwidth]{figures/figure4b}}\qquad
\subfloat{\includegraphics[width=0.3\textwidth]{figures/figure4c}}\\
\subfloat{\includegraphics[width=0.3\textwidth]{figures/figure4d}}\qquad
\subfloat{\includegraphics[width=0.3\textwidth]{figures/figure4e}}\qquad
\subfloat{\includegraphics[width=0.3\textwidth]{figures/figure4f}}
\end{flushleft}
\caption{The effect of the Gaussian $\aml$ prior on the posterior value of various model parameters. Results are plotted for the most probable systematic error model as given in, e.g., Table\,\ref{tab:all1}. 
The black line indicates a ratio of unity.}
\label{fig:alpha_multi}
\end{figure*}

\begin{figure*}
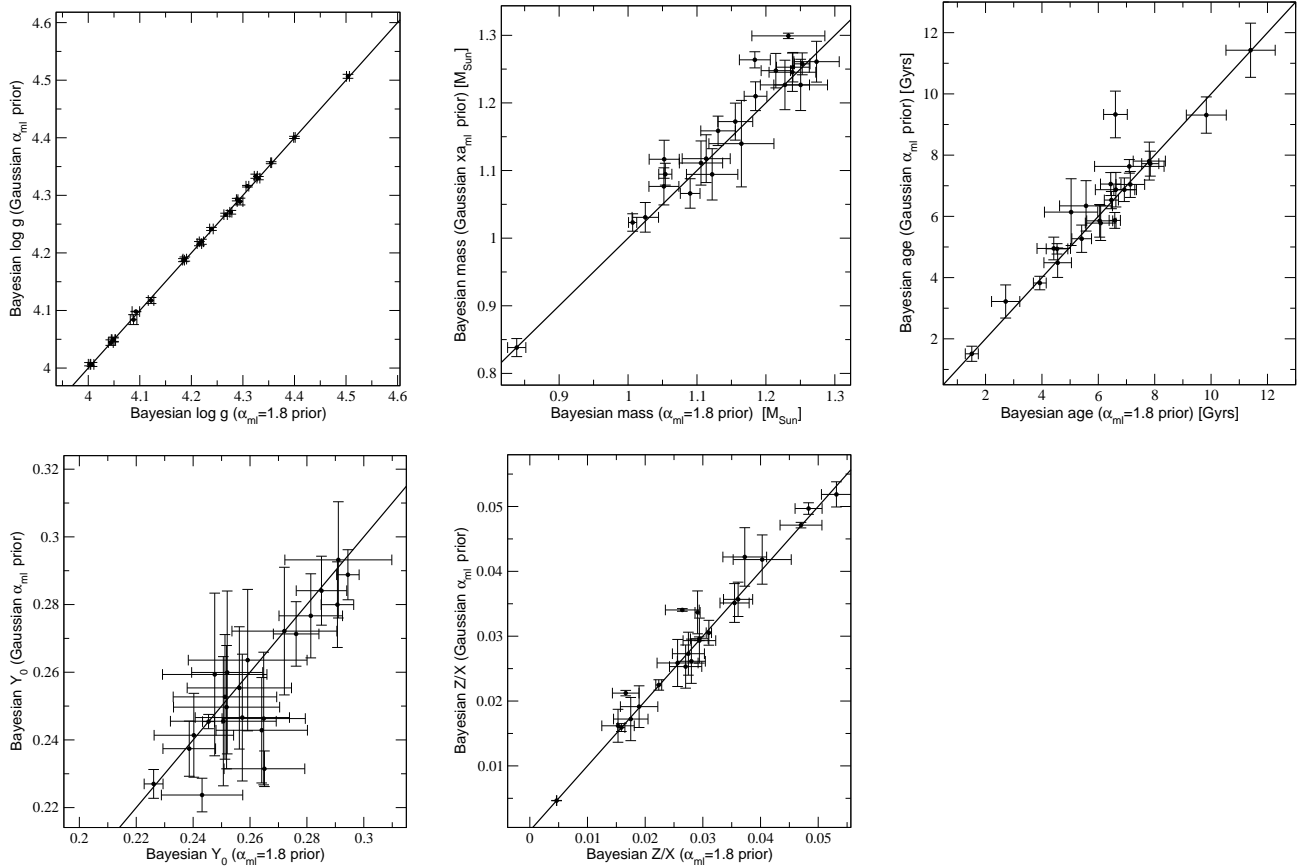

\begin{flushleft}
\subfloat{\includegraphics[width=0.3\textwidth]{figures/figure5a}}\qquad
\subfloat{\includegraphics[width=0.3\textwidth]{figures/figure5b}}\qquad
\subfloat{\includegraphics[width=0.3\textwidth]{figures/figure5c}}\\
\subfloat{\includegraphics[width=0.3\textwidth]{figures/figure5d}}\qquad
\subfloat{\includegraphics[width=0.3\textwidth]{figures/figure5e}}
\end{flushleft}
\caption{Same as Figure\,\ref{fig:alpha_multi} but for the $\aml = 1.8$ prior and the Gaussian $\aml$ prior. Comparison of $\aml$ is not shown.}
\label{fig:alpha_multi18}
\end{figure*}

Before we move on to a comparison to the literature, we first study the effect of the $\aml$ priors on our results. Figure\,\ref{fig:alpha_multi} shows the posterior mean values and uncertainties of $\aml$, $M$, $\YO$, $\log g$, $Z/X$, and age for all stars and compares the results with and without the Gaussian $\aml$ prior. The Gaussian $\aml$ prior leads to slightly lower values of $\aml$ as was expected from the discussion in Section\,\ref{sec:mixinglength}. Furthermore, the stellar masses are also slightly lower with an average difference $\left<\Delta M\right> = -0.021 \,\rm M_{\odot}$, and, although there is a larger scatter in $\YO$, slightly larger values in the initial helium mass fraction are also preferred with an average difference of $\left<\Delta\YO\right> =  0.008$. On the other hand, $\log g$ remains basically unaffected as expected, since the radius of the stars are well constrained by the large spacings (as we will see below, this also extends to a comparison with the other $\aml$ prior and the literature). $Z/X$ and age also do not show strong systematic effects. Nonetheless, the latter does exhibit a strong outlier with 16 Cyg B, for which the age changes from $9.279 \pm 0.473\,\rm Gyrs$ to $6.532 \pm 0.281\,\rm Gyrs$. Note that even though the Bayesian evidence is clearly in favour of the older model, the younger value is much more reasonable, given the results from \cite{metcalfe2012} and also given the value of the age for 16 Cyg A. This is a good test case for the impact of the $\aml$ prior.

A comparison of the results from the Gaussian prior and the fixed $\aml = 1.8$ prior is presented in Figure\,\ref{fig:alpha_multi18}. In general, the results fall in line with our expectations: the mass is now slightly larger for the Gaussian prior with $\left<\Delta M\right> = 0.014$, and $\YO$ is slightly smaller with $\left<\Delta\YO\right> = -0.005$. $Z/X$ and age values are again quite comparable except for a few outliers. In general the systematic differences between the Gaussian prior and the $\aml=1.8$ prior results are smaller than those obtained in a comparison without any priors on $\aml$. Overall, our comparison reveals that stronger constraints on $\aml$ do not perturb the parameters outside the uncertainties and  produce slightly lower stellar masses and higher $\YO$.

\begin{figure}
\centering
\includegraphics[width=\columnwidth]{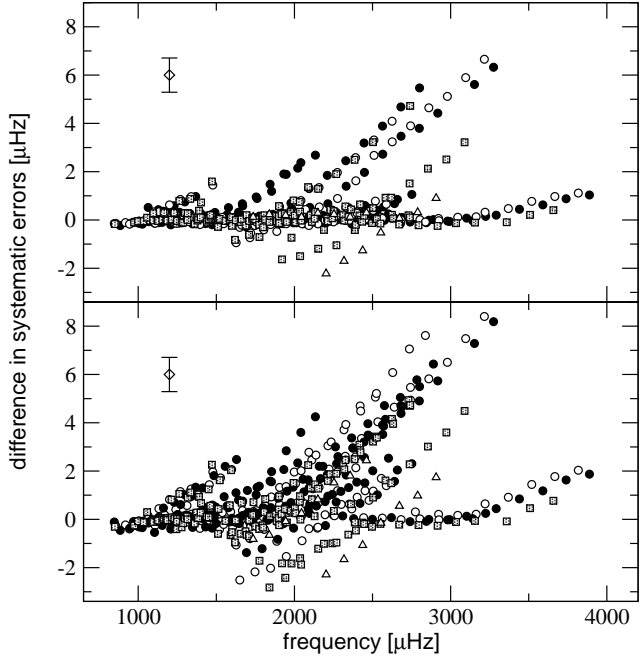}
\caption{Differences in the measured systematic errors that arise from using the Gaussian $\aml$ prior (top panel) or the $\aml = 1.8$ prior (bottom panel). All modes of all stars are shown: l=0 modes (open circles), l=1 modes (black circles), l=2 modes (shaded squares), and l=3 modes (open triangles). Positive (negative) values denote bigger (smaller) systematic errors in terms of surface effects when the $\aml$ priors are used. The average uncertainty of the differences is indicated by the diamond in the upper left. For each star, the plotted differences were obtained using the most probable systematic difference model for the respective $\aml$ prior.}
\label{fig:surfdiff}
\end{figure}

In terms of the systematic errors, in particular the surface effect, the results also follow our conclusions from the previous section. Figure\,\ref{fig:surfdiff} shows the differences in the systematic errors that arise by using the two $\aml$ priors for every mode of every star in our sample. Using the Gaussian $\aml$ prior leads to an increase in the magnitude of the surface effects (= more negative systematic differences between observed and calculated frequencies) in general. There are only a few stars in the sample for which the effect is very pronounced. It is interesting that for many modes the Gaussian $\aml$ prior does not produce large differences for the surface effects. This is due to the fact that we find a number of stars for which the surface effects are not significant unless we restrict the analysis to the $\aml=1.8$ models. Consequently, switching to the $\aml = 1.8$ prior results in even bigger surface effects and to significant surface effects for more stars in the sample. This is also reflected in the strong preference for the SSE model, as shown in the result tables. In both panels there are also a few outliers for which the priors lead to decreased surface effects (= more positive systematic differences between observed and calculated frequencies), but these modes belong to the stars for which the ASE model is either preferred or very similar in probability to the surface effect model.

As indicated in our result tables, using no $\aml$ prior often leads to the highest evidence. Larger evidence values require that the models are formally more consistent with all our available constraints while also minimizing the systematic errors, i.e., the surface effects. Therefore, the analysis which yields the highest evidence and thus the corresponding stellar parameters are usually interpreted as being most appropriate. As explained in Section\,\ref{sec:mixinglength}, however, we stress that at this point it is necessary to present the results from all approaches, and not to put too much confidence into the formal preference over to $\aml$ priors. This follows simply because we do not possess enough low-order modes or additional information to anchor the surface effect relation. The only clear exception to this are KIC 8379927 and KIC 10516096, for which we do not detect significant systematic errors irrespective of the $\aml$ priors but still find higher $\aml$ to be most probable. Concerning the impact of $\aml$ on the other stellar parameters, however, only the stellar mass and $\YO$ seem to be somewhat systematically affected by the choice of priors. Even for those parameters the deviations are usually within the quoted uncertainties. Thus, for our comparison with the values published in the literature, which also allow different values of $\aml$, we constrain ourselves to the results obtained using the ``intermediate approach", the Gaussian $\aml$ prior, and refer to our tabulated results for the differences arising from the different priors. 

\subsection{Surface effects and other systematic frequency differences}
\label{sec:evidence}

As previously alluded to, Figure\,\ref{fig:surfdiff} suggests that many stars do not show strong evidence for surface effects when our non-adiabatic models are used in tandem with the Gaussian prior. The situation changes, however, when the $\aml=1.8$ prior is used. This implies that, depending on the prior, the convective contributions to the surface effects are either more or less significant. Since a proper normalization of the surface effect amplitudes is not trivial and the shape of the surface effects can vary from star to star, we instead quantify the significance of the surface effect in terms of probabilities. As discussed in Section\,\ref{paper3:sec:fitting}, our calculations consider three different systematic error models: SSE, ASE, and NSE. Therefore, in order to quantify the surface effect significance for every star, we simply calculate
the odds ratio 
\begin{equation}
ODDS = \frac{ev(SSE)}{ ev(ASE) + ev(NSE)}, 
\end{equation}
where $ev(SSE)$, $ev(ASE)$, and $ev(NSE)$ are the evidence values obtained for the analysis using each specific systematic error model\footnote{This assumes that {\it a priori} all three surface effect models are equally probable.}.
This is the probability ratio between the hypotheses ``standard surface effect" and ``either arbitrary systematic errors or no systematic errors".
Therefore, if surface effects are needed to explain the observations, we expect that $ODDS \gg 1$. According to the convention established by \cite{jeffreys61}, the evidence for or against one of the two hypotheses is considered ``substantial" for a factor of 3 to 10, ``strong" for a factor of 10 to 30, ``very strong" for a factor of 30 to 100, and "decisive" for factors above 100. Hence, when the surface effects become more significant with respect to the other hypotheses, $ODDS$ will increase as well.

Our calculations show that for some stars the significance of the individual systematic error models depend on the specific prior for $\aml$, in accordance with what was discussed in Section\,\ref{sec:mixinglength}. However, there are four stars for which $ODDS<1$ irrespective of mixing length parameter: KIC 6933899, KIC 8379927, KIC 10516096, and Kepler-36. The latter three objects do not require any systematic errors at all. Furthermore, for KIC 6106415, KIC 6603624, and KIC 11244118 the surface effect model is only significant for the $\aml=1.8$ prior.

\begin{figure}
\centering
\includegraphics[width=\columnwidth]{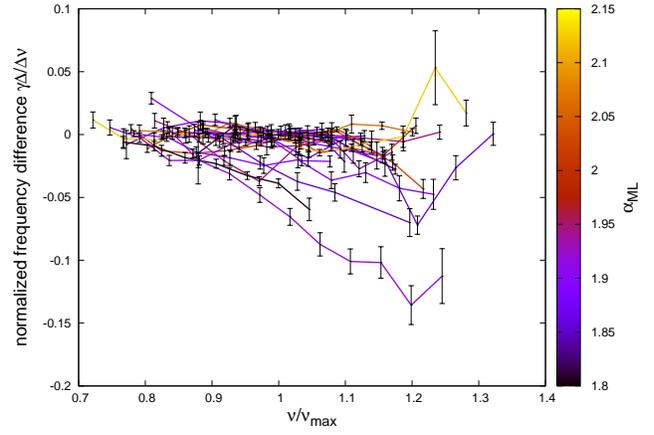}
\caption{Normalized systematic frequency differences as a function of normalized frequency for $l=0$ modes for the results obtained with the Gaussian $\aml$ prior. The colour represents the mean posterior $\aml$. For each star, the plotted differences were obtained using the most probable systematic difference model.}
\label{fig:surfmeasure}
\end{figure}

Figure\,\ref{fig:surfmeasure} shows the actual systematic error measurements obtained when using the Gaussian $\aml$ prior that have been rescaled and plotted as a function of their mean $\aml$.
For many stars the individual deviations do not seem to correspond to the clear power-law behaviour that can be identified for the Sun. Furthermore, there appears to be a very weak dependence on $\aml$, where higher values are related to smaller normalised surface effects, as expected from the discussion in Section\,\ref{sec:mixinglength}. Whether this dependence is physically meaningful depends on whether these stars actually have higher values of $\aml$, or if it is simply the case that our $\aml$ prior is too weak. In any case, $\aml$ and surface effects are related. 

\begin{figure}
\centering
\includegraphics[width=\columnwidth]{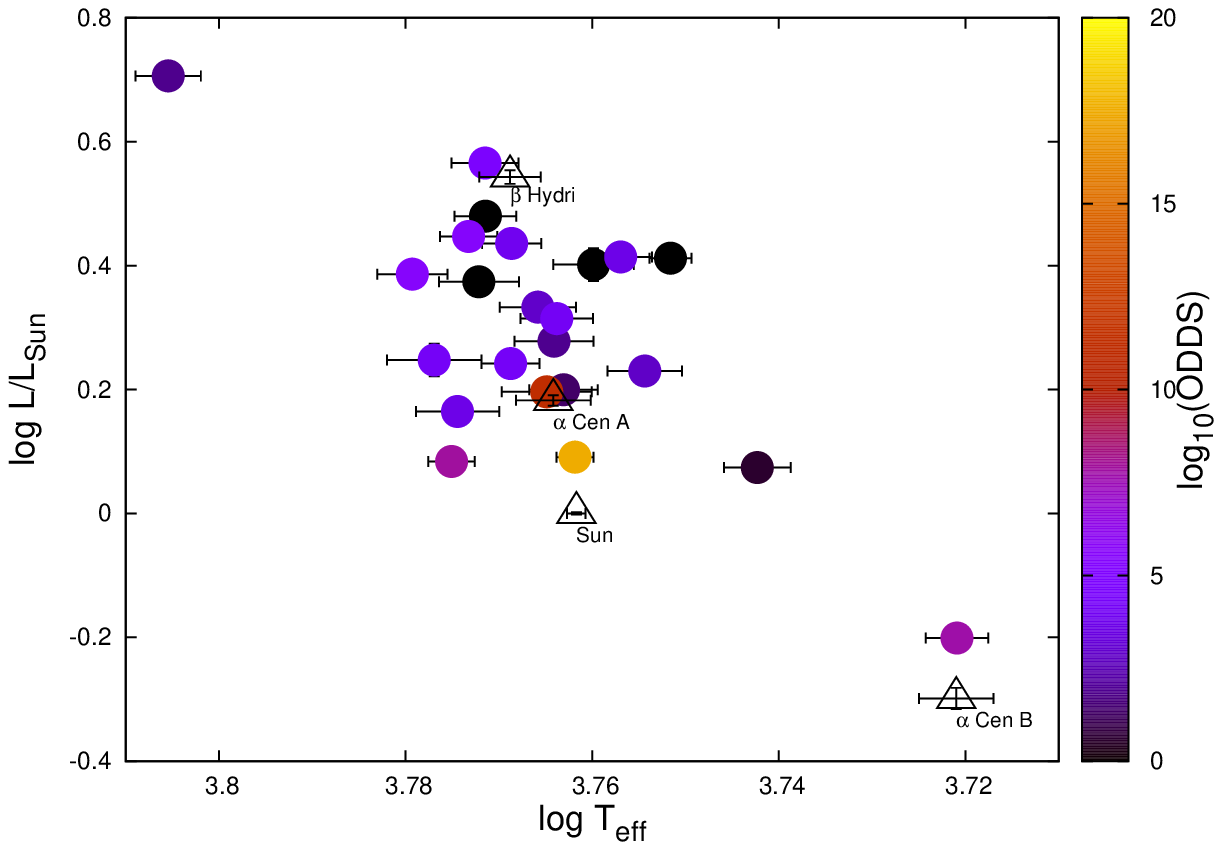}
\includegraphics[width=\columnwidth]{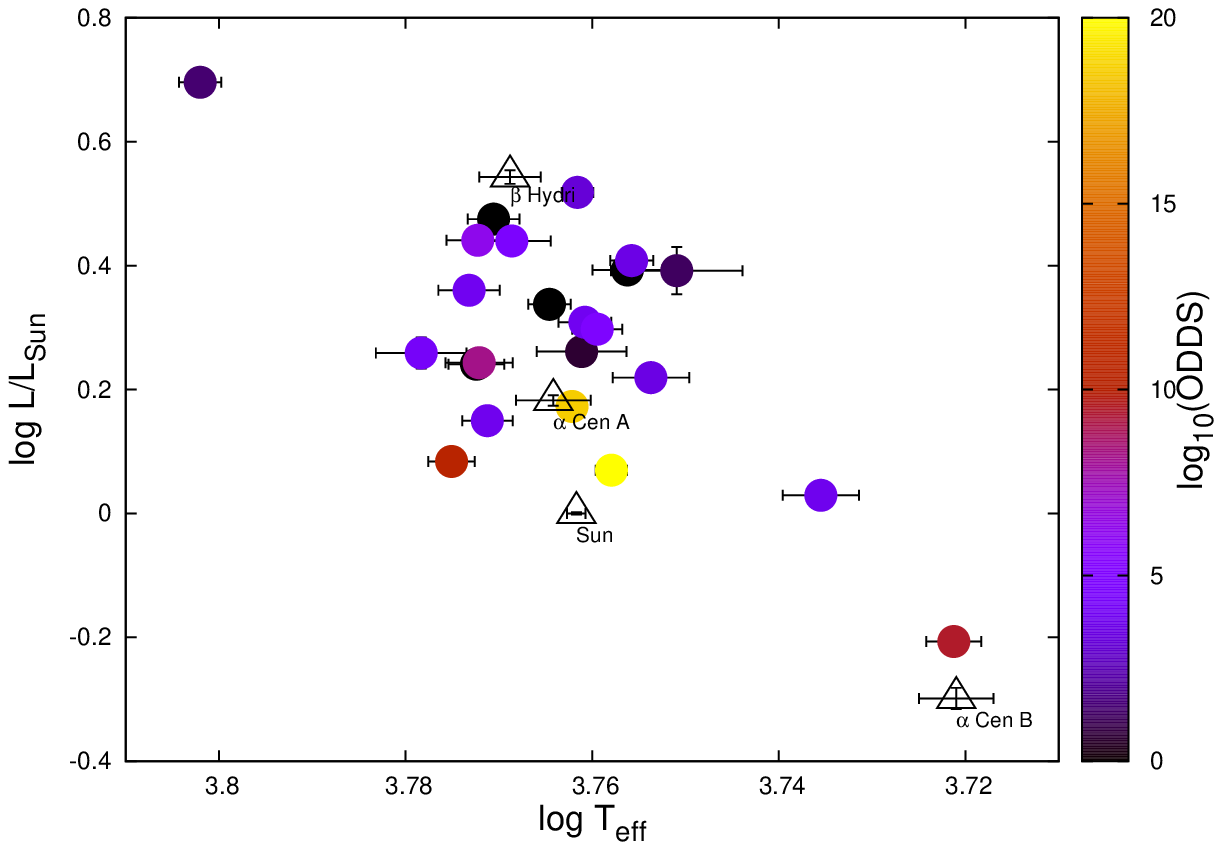}
\caption{HR diagram of all stars in our sample (filled circles) with parameter taken from Table\,\ref{tab:all1} to Table\,\ref{tab:all3}, using the results from the Gaussian (top panel) and $\aml=1.8$ (bottom panel) prior. The colour indicates the significance of the detected surface effect using $\log_{10}(ODDS)$. Four other well-known stars with surface effects are also shown as triangles. For each star, the plotted parameters were obtained using the most probable systematic difference model for the respective $\aml$ prior.}
\label{fig:curious}
\end{figure}

Similar to \cite{mathur2012}, we do not find any simple correlations of the normalised surface effect with any of the other parameters in Table\,\ref{tab:all1} to Table\,\ref{tab:all3}. However, studying the {\it significance} of the surface effect in terms of probabilities reveals some interesting results. Figure\,\ref{fig:curious} shows the logarithm of the odds ratio for all stars in our sample as a function of their position in the HR diagram. The most significant detections appear to be situated at close-to-solar values of $T_{\rm eff}$ and the picture is similar whether the Gaussian $\aml$ prior, the $\aml = 1.8$ prior, or no $\aml$ prior is used. Furthermore, the coolest star in the sample, KIC 8006161, also displays highly significant surface effects but lies far off from the main bulk of the sample. We have also added symbols representing the Sun \citep{gruberbauer2013}, $\beta$ Hydri \citep{brandao2011}, and $\alpha$ Cen A\& B \citep{eggenberger2004}, all of which were used by \cite{kjeldsen2008} to define the surface effect correction. Except for $\beta$ Hydri\footnote{Note that the surface effects detected in $\beta$ Hydri have only been measured using adiabatic frequencies which do not contain the correction for radiative gains and losses.}, the stars fit well into the pattern given by the Kepler stars. 16 Cyg A, 16 Cyg B,$\alpha$ Cen A, and of course the Sun, appear to lie on the ``surface effect locus" in the HRD diagram of our sample. $\alpha$ Cen B, on the other hand, is situated very close to KIC 800616. 

The stars for which no significant surface effects were detected do mix with stars that show less significant detections, which is why there does not seem to be a strong correlation of the surface effect with any particular parameter.
On average, however, lower luminosities and higher effective temperatures correspond to more significant surface effects. Plotting $\log T_{\rm eff}$ against $\log g$ (not shown) necessarily yields a very similar picture which again clusters the most significant detections at the solar values. A correlation of the surface effect amplitude with $\log g$ was already noted by \cite{mathur2012}. Our comparison of the significance of the surface effect would be more in line with their investigation of the normalised surface effect for which they could not find a strong correlation. It will be intriguing to see whether a bigger sample and additional lower order modes could lead to a clearer detection of a ``surface effect locus" in the HR diagram.

In any case, the non-detection of surface effects in some stars, as well as the concentration of very significant surface effects for stars with close to solar values should be a warning for 
unreflected usage of the standard surface correction for all solar-like stars. 

\subsection{Comparison with non-Bayesian results}

\subsubsection{Mathur20}
\label{sec:amp}

\begin{figure*}
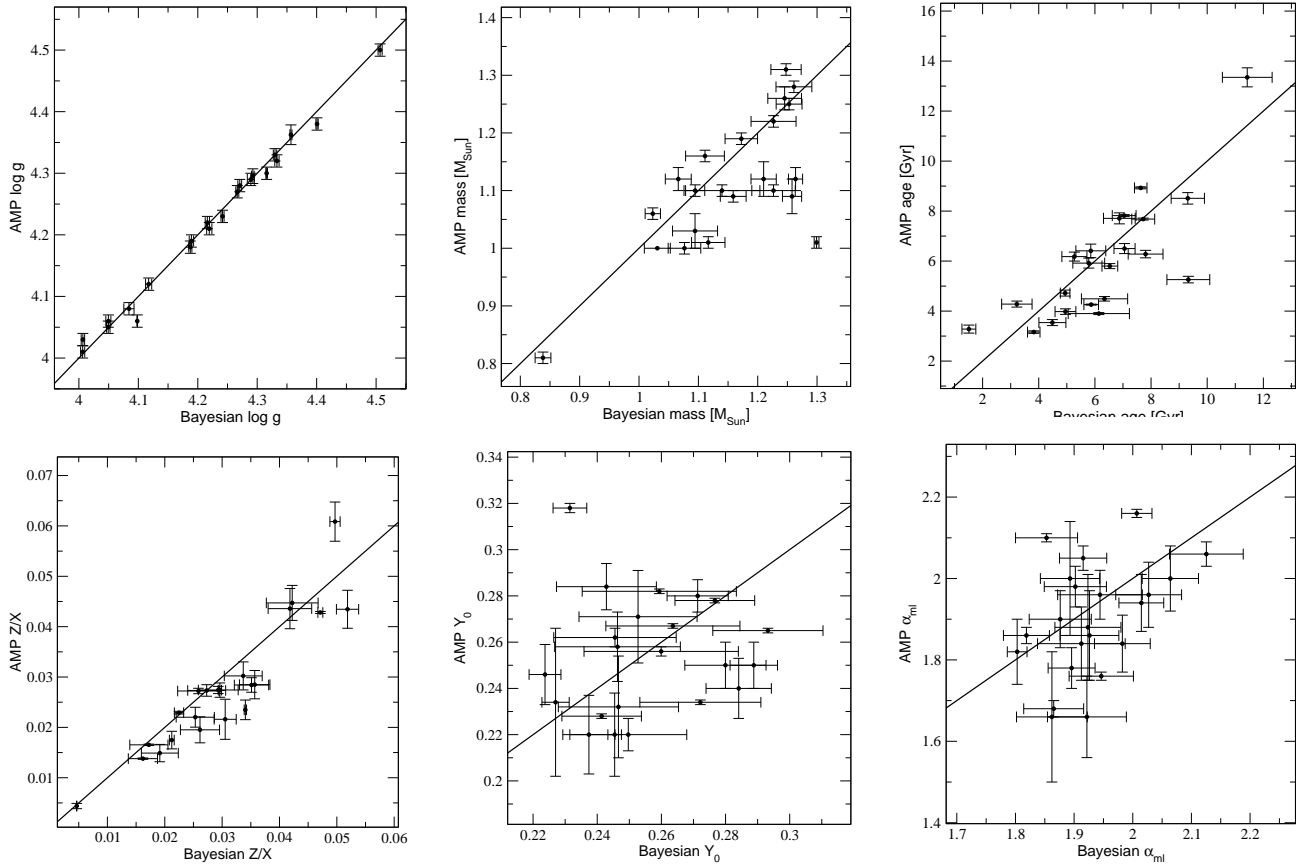

\begin{flushleft}
\subfloat{\includegraphics[width=0.3\textwidth]{figures/figure9a}}\qquad
\subfloat{\includegraphics[width=0.3\textwidth]{figures/figure9b}}\qquad
\subfloat{\includegraphics[width=0.3\textwidth]{figures/figure9c}}\\
\subfloat{\includegraphics[width=0.3\textwidth]{figures/figure9d}}\qquad
\subfloat{\includegraphics[width=0.3\textwidth]{figures/figure9e}}\qquad
\subfloat{\includegraphics[width=0.3\textwidth]{figures/figure9f}}
\end{flushleft}
\caption{Same as Figure\,\ref{fig:alpha_multi} but comparing the results from our Bayesian approach using the $\aml$ prior with the published results obtained via the AMP pipeline \citep{mathur2012, metcalfe2012}. Note that Kepler-36 is not included in these plots.}
\label{fig:mathur_comp}
\end{figure*}

In this section we investigate the presence of potential systematic differences between our results and those obtained using the AMP pipeline. Figure\,\ref{fig:mathur_comp} shows that there are no strong systematic trends in either of the plotted parameters. As in the comparison between our three different Bayesian analyses (Figure\,\ref{fig:alpha_multi}, Figure\,\ref{fig:alpha_multi18}), the determined $\log g$ values are very similar, but the Bayesian uncertainties are usually smaller. The results for $\aml$ show large scatter which is mostly compensated by the large uncertainties. It should be noted that our grid only extends from $\aml = 1.8$ to 2.4, and therefore we do not cover the lower values that AMP returns for some of the stars. 

The masses that were determined are quite similar for most stars, but the AMP delivers smaller uncertainties on average. For several stars, only the larger uncertainties reported by the Bayesian method can help to reconcile the results. There exists also a clear outlier with KIC 11244118 where the masses differ by about $0.3\,M_{\odot}$, more than 15 times our statistical uncertainty.\footnote{This star is also problematic since it fits best to models near the border of our grid both in terms of mass and metallicity.} The initial helium mass fraction again displays large but seemingly unsystematic scatter, in particular when compared to some of the uncertainties reported by AMP. Many of the stars appear to prefer very low values of $\YO$, as was also found to be the case by \cite{mathur2012}. However, for these stars our various approaches (different $\aml$ priors, different systematic error models) can often provide a solution with higher values albeit lower evidence. Also, in many cases the Bayesian uncertainties are usually large enough to reconcile the values with those required from studies of Big Bang nucleosynthesis. The only clear outlier here is KIC 8379927 for which we find quite large disagreements with the AMP results. Contrary to the somewhat larger discrepancies for $\YO$, the results for $Z/X$ are more similar, but our values appear to be slightly larger in a systematic way. In general, we have to stress that concerning the chemical composition, our grid is quite coarse compared to the capabilities of AMP's genetic algorithm. 

Lastly, significant differences appear in the comparison of the determined ages. Irrespective of potential differences in the definition of zero-age models, the two methods yield different results with significant scatter. Moreover, the Bayesian age uncertainties appear to be bigger on average by a factor of 6, which is substantial, necessary, but insufficient to reconcile the results in many cases. 

We re-emphazise that the Bayesian uncertainties are properly propagated through the whole grid and also include the effects of the systematic frequency differences (via marginalisation) and any non-asteroseismic constraints (via the prior probabilities). AMP, on the other hand, can only consider statistical contributions to the uncertainties. While dependent on the particular grid that was analysed, the Bayesian uncertainties are therefore superior from a methodological point of view. This different approach, as well as differences in the stellar models themselves, is sufficient to explain the reported discrepancies.  

\subsubsection{16 Cyg A \& B}

The modelling performed by \cite{metcalfe2012} revealed that 16 Cyg A \& B are of slightly different masses but have a similar age, as 
expected for a binary system. Several different grids and methods were used, including AMP, to arrive at an average ensemble solution. Our results
compare favourably with this ensemble average, when it comes to the ages, the masses, $\ZO$, and $\aml$. Except for the mass of 16 Cyg B, for which we obtain $1.023 \pm 0.013\,M_{\odot}$ compared to their result of $1.07 \pm 0.02\,M_{\odot}$, these parameters overlap within their respective $1\sigma$ uncertainties. It should be noted that we obtain a lower mass for 16 Cyg A as well, which might suggest a systematic difference between the methods and models used. As discussed in the previous section, however, we don't find that such a trend is true for our larger sample. The ages are fully consistent with a common origin, even though this constraint was not used in the analysis\footnote{The equal age is in even better agreement with our results for the $\aml=1.8$ prior, but for this approach we also obtain substantially smaller masses.}.

We find a slight discrepancy for the initial helium mass fraction. For 16 Cyg A we obtain $\YO = 0.282 \pm 0.01$ and for 16 Cyg B we find $\YO = 0.285 \pm 0.01$, while \citeauthor{metcalfe2012} report $0.25 \pm 0.01$. Overall, we observe that the differences between our results and the ensemble average in the literature are minor. 

Comparing our results exclusively to the AMP values, we see a significant difference in the age and the value of $\aml$ for 16 Cyg B. It is interesting that this star is among the set of the most significant surface-effect detections in our sample. As the AMP results in a value of $\aml = 2.05 \pm 0.03$, which is bigger than the ensemble average, it is perhaps the combination of the solar-calibrated surface effect correction and the use of a higher-than-solar $\aml$ which results in the discrepancy. For the age, we obtained $6.532 \pm 0.281\,\rm Gyr$ compared to $5.8 \pm 0.1$. Consistent with our findings in Section\,\ref{sec:amp}, we observe that our age uncertainties are significantly bigger.

In a recent paper, \cite{white2013} have combined interferometric diameters from CHARA observations with Hipparcos parallaxes, spectrophotometric bolometric fluxes, and the asteroseismic large frequency separation, to obtain largely model-independent constraints for 16 Cyg A \& B. In comparison to their results, for 16 Cyg A, our $\aml = 1.8$ prior produces a very close match in terms of mass and radius, but the model $T_{\rm eff}$ values are slightly too low and match better for the Gaussian $\aml$ prior. For 16 Cyg B, on the other hand, the higher $\aml$ values are more consistent with their results, predicting higher masses and larger radii but again $T_{\rm eff}$ values that are not quite high enough to match the mean observed values. These slight differences however are insignificant and, irrespective of the particular priors used, we find that our results match the masses, temperatures, and radii from \cite{white2013} reasonably well and in all cases to within the combined 1.5$\sigma$ uncertainties. Therefore, the interferometric uncertainties are too large to give strong evidence for or against our particular solutions (i.e., in particular the different $\aml$ values). This can also be interpreted as additional justification for the various $\aml$ priors, since the range of results allows us to define a parameter space that is more in line with model-independent observations. 

\subsubsection{Kepler-36}

With respect to Kepler-36, we find that we can match all parameters published in \cite{carter2012} within the uncertainties. It is interesting, however, that we do not detect any surface effects for this star. \citeauthor{carter2012} report that the surface-effect correction was applied to the frequencies. Judging from our results, any surface effects necessary to be corrected for this star would have to originate from the radiative losses that are already taken into account in our non-adiabatic models.

\section{Conclusions}

In this paper we have reported on our asteroseismic analysis of 23 previously published stars that were observed with the Kepler satellite.
We compared the results obtained with our Bayesian grid-based method to the results from the literature, most importantly those obtained with the AMP. Except for a weak trend towards larger values of $Z/X$ with our method, no obvious systematic differences in the basic stellar parameters can be found. In part, this is certainly due to spectroscopic constraints ($T_{\rm eff}$, $\log g$, $[Fe/H]$, $L/L_{\odot}$) that were used by all authors.
 
However, we observe that the uncertainties derived from the two methods differ substantially for some stellar parameters. Uncertainties in the stellar ages in particular are either significantly underestimated by AMP or significantly overestimated by the Bayesian method. We conclude that the flexible treatment of the surface effects in the Bayesian approach is probably responsible for this discrepancy. Different values of $\aml$ and the usage of non-adiabatic models require a more flexible treatment of the surface effect. Therefore, in our view the uncertainties derived with our method more adequately represent our actual state of knowledge about the surface effects and are therefore more realistic. On the other hand, the interplay between the surface effect and $\aml$ introduces another layer of complexity in the analysis which has to be taken into account in the determination of the stellar parameters. We propose that future studies with more stars should aim to reexamine this interdependence, especially as long as non-seismic constraints on $\aml$ are not available. 

Concerning the surface effects themselves, we find that with a Gaussian prior on $\aml$, only a few stars in our sample actually require larger corrections. 6 stars in our sample do not show strong evidence for any surface effect at all. Compared to the results in \cite{mathur2012}, this suggests that for many stars taking into account the radiative losses is already good enough. On the other hand, using only models with $\aml = 1.8$ leads to more significant detections. Irrespective of the prior on $\aml$, we also discovered that the stars for which we do find a highly significant surface effect appear to be located very close to the Sun in the HR diagram (see Figure\,\ref{fig:curious}). A comparison with the stars that were used to derive the traditional surface-effect correction \citep{kjeldsen2008} shows that most of these calibrators - including the Sun - also fit the picture. As radiative losses are already taken into account in our models, the modelling of convection and its dependencies on element abundances, opacities, and the equation of state remains a leading candidate to explain the cause of the surface effects.

To conclude, although systematic differences between stellar evolutionary codes are still affecting the individual stellar parameters, the systematic analysis of surface effects can already be pursued using more advanced methods than the standard surface correction, such as our Bayesian approach. No matter which surface correction is used, however, the constraints on $\aml$ will potentially affect the results in the absence of lower-order modes. 
The data sets on which this analysis is based have since been superseded by many more quarters of Kepler data. Also, many more stars have been observed, for which public frequencies are also available \citep{appourchaux2012}. Strong spectroscopic constraints and access to lower-order modes will be necessary to improve our analysis, and to see whether the ``surface effect locus" can be reproduced with a larger sample of stars and better data. Given the large number of subgiants and red giants observed with {\it Kepler} and CoRoT, a similar study for non-main sequence stars could be very illuminating as well.

\section*{Acknowledgments}

MG, DG, and KM acknowledge funding from the Natural Sciences and Engineering Research Council of Canada. Computational facilities were provided by ACEnet, which is funded by the Canada Foundation for Innovation (CFI),  Atlantic Canada Opportunities Agency (ACOA), and the provinces of Newfoundland and Labrador, Nova Scotia, and New Brunswick. TK acknowledges financial support form the Austrian Science Fund (FWF P23608).
 
\bibliography{starpaper}

\begin{table*}
   \centering
   \caption{Mean parameters and uncertainties as a function of $\alpha_{\rm ml}$ prior for KIC 3632418 to KIC 6603624. Bold font indicates the prior for which the highest evidence was obtained, as well as other priors for which the evidence was comparable (within a factor of 5). $\YO$, $\ZO$: initial helium and metal mass fractions; $\Zs$: metal mass fraction in the envelope; $R_{\rm BCZ}$: fractional radius of the base of the convection zone; $\alpha_{\rm ml}$: mixing length parameter; sys: the most probable systematic-error model is given (SSE = standard surface effect, ASE = arbitrary systematic errors, NSE = no systematic errors) and asterisks indicate a probability contrast of less than an order of magnitude with respect to any of the other systematic-error models.}
   \resizebox{0.99\textwidth}{!} {
     \begin{tabular}{l l c c c c c c c c c c c l} 
   \hline
   \hline
     Star & $\alpha_{\rm}$ prior & $M/M_{\odot}$ & $\log T_{\rm eff}$ & $\log L/L_{\odot}$ & $\log R/R_{\odot}$ & Age & $\YO$ & $\ZO$ & $\Zs$ & $\ZoX$ & $R_{\rm BCZ}$ & $\alpha_{\rm ml}$ & sys \\
     \hline
3632418 &  $\bm{\alpha_{\rm ml}=1.8}$   & $1.273$  & $3.802$  & $0.696$  & $0.268$  & $3.926$  & $0.252$  & $0.0134$  & $0.0130$  & $0.0175$  & $0.8397$  & $1.80$  & SSE  \\ 
& & $ \pm 0.033$  & $ \pm 0.002$  & $ \pm 0.007$  & $ \pm 0.004$  & $ \pm 0.227$  & $ \pm 0.012$  & $ \pm 0.0024$  & $ \pm 0.0022$  & $ \pm 0.0030$  & $ \pm 0.0069$  & & \\
& \textbf{Gaussian} & $1.261$  & $3.805$  & $0.706$  & $0.266$  & $3.823$  & $0.260$  & $0.0130$  & $0.0126$  & $0.0172$  & $0.8405$  & $1.87$  & SSE  \\ 
& & $ \pm 0.030$  & $ \pm 0.003$  & $ \pm 0.012$  & $ \pm 0.004$  & $ \pm 0.221$  & $ \pm 0.024$  & $ \pm 0.0025$  & $ \pm 0.0022$  & $ \pm 0.0033$  & $ \pm 0.0088$  & $ \pm 0.05$  & \\ 
& \textbf{no} $\bm\aml$ \textbf{prior} & $1.264$  & $3.807$  & $0.713$  & $0.266$  & $3.738$  & $0.264$  & $0.0133$  & $0.0129$  & $0.0177$  & $0.8386$  & $1.91$  & SSE  \\ 
& & $ \pm 0.029$  & $ \pm 0.003$  & $ \pm 0.012$  & $ \pm 0.004$  & $ \pm 0.217$  & $ \pm 0.024$  & $ \pm 0.0024$  & $ \pm 0.0021$  & $ \pm 0.0032$  & $ \pm 0.0093$  & $ \pm 0.06$  & \\
\hline
3656476 & $\alpha_{\rm ml}=1.8$ & $1.131$  & $3.754$  & $0.219$  & $0.126$  & $6.623$  & $0.281$  & $0.0310$  & $0.0273$  & $0.0373$  & $0.6874$  & $1.80$  & SSE  \\ 
 & & $ \pm 0.025$  & $ \pm 0.004$  & $ \pm 0.017$  & $ \pm 0.003$  & $ \pm 0.729$  & $ \pm 0.011$  & $ \pm 0.0028$  & $ \pm 0.0026$  & $ \pm 0.0038$  & $ \pm 0.0090$  &  & \\ 
 & Gaussian & $1.159$  & $3.754$  & $0.230$  & $0.130$  & $6.871$  & $0.276$  & $0.0347$  & $0.0308$  & $0.0422$  & $0.6732$  & $1.94$ & SSE  \\ 
 & & $ \pm 0.022$  & $ \pm 0.004$  & $ \pm 0.017$  & $ \pm 0.003$  & $ \pm 0.564$  & $ \pm 0.012$  & $ \pm 0.0034$  & $ \pm 0.0031$  & $ \pm 0.0045$  & $ \pm 0.0089$  & $ \pm 0.07$  & \\ 
 & \textbf{no} $\bm\aml$ \textbf{prior} & $1.253$  & $3.766$  & $0.301$  & $0.143$  & $7.789$  & $0.234$  & $0.0400$  & $0.0359$  & $0.0473$  & $0.6591$  & $2.39$  & NSE*  \\ 
 & & $ \pm 0.013$  & $ \pm 0.002$  & $ \pm 0.007$  & $ \pm 0.002$  & $ \pm 0.287$  & $ \pm 0.010$  & $ \pm 0.0005$  & $ \pm 0.0005$  & $ \pm 0.0010$  & $ \pm 0.0015$  & $ \pm 0.02$  & \\ 
\hline
4914923 & $\alpha_{\rm ml}=1.8$& $1.228$  & $3.759$  & $0.297$  & $0.154$  & $5.409$  & $0.259$  & $0.0306$  & $0.0271$  & $0.0361$  & $0.7097$  & $1.80$  & SSE  \\ 
 & & $ \pm 0.036$  & $ \pm 0.003$  & $ \pm 0.013$  & $ \pm 0.004$  & $ \pm 0.349$  & $ \pm 0.021$  & $ \pm 0.0017$  & $ \pm 0.0016$  & $ \pm 0.0025$  & $ \pm 0.0078$  & & \\ 
 & \textbf{Gaussian} & $1.227$  & $3.764$  & $0.314$  & $0.153$  & $5.269$  & $0.263$  & $0.0299$  & $0.0266$  & $0.0357$  & $0.7075$  & $1.88$  & SSE  \\ 
 & & $ \pm 0.037$  & $ \pm 0.004$  & $ \pm 0.018$  & $ \pm 0.004$  & $ \pm 0.446$  & $ \pm 0.021$  & $ \pm 0.0018$  & $ \pm 0.0017$  & $ \pm 0.0026$  & $ \pm 0.0092$  & $ \pm 0.05$  & \\ 
 & \textbf{no} $\bm\aml$ \textbf{prior} &  $1.245$  & $3.769$  & $0.343$  & $0.157$  & $6.802$  & $0.242$  & $0.0337$  & $0.0302$  & $0.0399$  & $0.6812$  & $2.19$  & SSE  \\ 
 & & $ \pm 0.025$  & $ \pm 0.005$  & $ \pm 0.021$  & $ \pm 0.003$  & $ \pm 0.766$  & $ \pm 0.020$  & $ \pm 0.0035$  & $ \pm 0.0032$  & $ \pm 0.0044$  & $ \pm 0.0102$  & $ \pm 0.12$  & \\ 
\hline
5184732 & $\alpha_{\rm ml}=1.8$& $1.239$  & $3.761$  & $0.261$  & $0.132$  & $4.421$  & $0.277$  & $0.0394$  & $0.0350$  & $0.0483$  & $0.7258$  & $1.80$  & SSE*  \\ 
 & & $ \pm 0.024$  & $ \pm 0.005$  & $ \pm 0.023$  & $ \pm 0.003$  & $ \pm 0.594$  & $ \pm 0.008$  & $ \pm 0.0016$  & $ \pm 0.0015$  & $ \pm 0.0023$  & $ \pm 0.0143$  & & \\ 
 & Gaussian & $1.253$  & $3.764$  & $0.278$  & $0.135$  & $4.951$  & $0.271$  & $0.0400$  & $0.0360$  & $0.0497$  & $0.6995$  & $2.03$  & SSE  \\ 
 & & $ \pm 0.022$  & $ \pm 0.004$  & $ \pm 0.020$  & $ \pm 0.003$  & $ \pm 0.370$  & $ \pm 0.010$  & $ \pm 0.0004$  & $ \pm 0.0004$  & $ \pm 0.0009$  & $ \pm 0.0070$  & $ \pm 0.06$  & \\ 
 & \textbf{no} $\bm\aml$ \textbf{prior} & $1.274$  & $3.771$  & $0.312$  & $0.137$  & $4.521$  & $0.273$  & $0.0399$  & $0.0362$  & $0.0502$  & $0.7022$  & $2.15$  & SSE  \\ 
 & & $ \pm 0.016$  & $ \pm 0.003$  & $ \pm 0.016$  & $ \pm 0.002$  & $ \pm 0.257$  & $ \pm 0.008$  & $ \pm 0.0007$  & $ \pm 0.0006$  & $ \pm 0.0011$  & $ \pm 0.0049$  & $ \pm 0.05$  & \\ 
\hline
5512589 & $\bm{\aml=1.8}$& $1.106$  & $3.756$  & $0.408$  & $0.216$  & $7.843$  & $0.272$  & $0.0222$  & $0.0192$  & $0.0256$  & $0.6620$  & $1.80$  & SSE  \\ 
 & & $ \pm 0.031$  & $ \pm 0.002$  & $ \pm 0.008$  & $ \pm 0.004$  & $ \pm 0.303$  & $ \pm 0.018$  & $ \pm 0.0026$  & $ \pm 0.0024$  & $ \pm 0.0036$  & $ \pm 0.0051$  & & \\ 
 & \textbf{Gaussian} & $1.111$  & $3.757$  & $0.414$  & $0.217$  & $7.722$  & $0.272$  & $0.0223$  & $0.0194$  & $0.0259$  & $0.6629$  & $1.82$  & SSE  \\ 
 & & $ \pm 0.033$  & $ \pm 0.003$  & $ \pm 0.015$  & $ \pm 0.004$  & $ \pm 0.408$  & $ \pm 0.019$  & $ \pm 0.0026$  & $ \pm 0.0024$  & $ \pm 0.0036$  & $ \pm 0.0054$  & $ \pm 0.04$  & \\ 
 & \textbf{no} $\bm\aml$ \textbf{prior} & $1.117$  & $3.758$  & $0.421$  & $0.218$  & $7.588$  & $0.272$  & $0.0225$  & $0.0196$  & $0.0261$  & $0.6640$  & $1.84$  & SSE  \\ 
 & & $ \pm 0.034$  & $ \pm 0.004$  & $ \pm 0.019$  & $ \pm 0.004$  & $ \pm 0.472$  & $ \pm 0.019$  & $ \pm 0.0026$  & $ \pm 0.0025$  & $ \pm 0.0037$  & $ \pm 0.0057$  & $ \pm 0.05$  & \\ 
\hline
6106415 & $\alpha_{\rm ml}=1.8$& $1.184$  & $3.772$  & $0.243$  & $0.101$  & $4.536$  & $0.243$  & $0.0236$  & $0.0204$  & $0.0264$  & $0.7446$  & $1.80$  & SSE  \\ 
 & & $ \pm 0.022$  & $ \pm 0.004$  & $ \pm 0.014$  & $ \pm 0.003$  & $ \pm 0.383$  & $ \pm 0.014$  & $ \pm 0.0023$  & $ \pm 0.0021$  & $ \pm 0.0029$  & $ \pm 0.0088$  & & \\ 
 & Gaussian & $1.264$  & $3.772$  & $0.264$  & $0.112$  & $4.939$  & $0.224$  & $0.0300$  & $0.0265$  & $0.0341$  & $0.7174$  & $2.06$  & NSE  \\ 
 & & $ \pm 0.012$  & $ \pm 0.002$  & $ \pm 0.010$  & $ \pm 0.001$  & $ \pm 0.170$  & $ \pm 0.005$  & $ \pm < 0.0001$  & $ \pm 0.0001$  & $ \pm 0.0002$  & $ \pm 0.0038$  & $ \pm 0.05$  & \\ 
 & \textbf{no} $\bm\aml$ \textbf{prior} & $1.267$  & $3.774$  & $0.271$  & $0.112$  & $4.922$  & $0.223$  & $0.0299$  & $0.0265$  & $0.0340$  & $0.7163$  & $2.10$  & NSE*  \\ 
 & & $ \pm 0.007$  & $ \pm 0.002$  & $ \pm 0.008$  & $ \pm 0.001$  & $ \pm 0.150$  & $ \pm 0.005$  & $ \pm 0.0008$  & $ \pm 0.0007$  & $ \pm 0.0009$  & $ \pm 0.0029$  & $ \pm 0.02$  & \\ 
\hline
6116048 & $\alpha_{\rm ml}=1.8$& $1.090$  & $3.772$  & $0.241$  & $0.099$  & $6.608$  & $0.239$  & $0.0159$  & $0.0132$  & $0.0166$  & $0.7290$  & $1.80$  & ASE  \\ 
 & & $ \pm 0.014$  & $ \pm 0.003$  & $ \pm 0.011$  & $ \pm 0.002$  & $ \pm 0.420$  & $ \pm 0.009$  & $ \pm 0.0019$  & $ \pm 0.0017$  & $ \pm 0.0023$  & $ \pm 0.0073$  & & \\ 
 & Gaussian & $1.066$  & $3.763$  & $0.200$  & $0.097$  & $9.328$  & $0.237$  & $0.0200$  & $0.0167$  & $0.0212$  & $0.6747$  & $2.01$  & SSE  \\ 
 & & $ \pm 0.022$  & $ \pm 0.004$  & $ \pm 0.019$  & $ \pm 0.003$  & $ \pm 0.763$  & $ \pm 0.008$  & $ \pm 0.0003$  & $ \pm 0.0003$  & $ \pm 0.0004$  & $ \pm 0.0082$  & $ \pm 0.04$  & \\ 
 & \textbf{no} $\bm\aml$ \textbf{prior} & $1.082$  & $3.770$  & $0.230$  & $0.099$  & $8.650$  & $0.238$  & $0.0197$  & $0.0166$  & $0.0212$  & $0.6789$  & $2.12$  & SSE  \\ 
 & & $ \pm 0.027$  & $ \pm 0.005$  & $ \pm 0.024$  & $ \pm 0.004$  & $ \pm 0.865$  & $ \pm 0.009$  & $ \pm 0.0012$  & $ \pm 0.0010$  & $ \pm 0.0013$  & $ \pm 0.0100$  & $ \pm 0.06$  & \\ 
\hline
6603624 & $\alpha_{\rm ml}=1.8$& $1.052$  & $3.735$  & $0.029$  & $0.067$  & $9.830$  & $0.264$  & $0.0356$  & $0.0301$  & $0.0403$  & $0.6625$  & $1.80$  & SSE  \\ 
 & & $ \pm 0.022$  & $ \pm 0.004$  & $ \pm 0.015$  & $ \pm 0.003$  & $ \pm 0.708$  & $ \pm 0.016$  & $ \pm 0.0040$  & $ \pm 0.0034$  & $ \pm 0.0050$  & $ \pm 0.0057$  & & \\ 
 & Gaussian & $1.117$  & $3.742$  & $0.074$  & $0.076$  & $9.309$  & $0.243$  & $0.0373$  & $0.0319$  & $0.0418$  & $0.6627$  & $1.98$  & SSE*  \\ 
 & & $ \pm 0.028$  & $ \pm 0.004$  & $ \pm 0.017$  & $ \pm 0.004$  & $ \pm 0.593$  & $ \pm 0.015$  & $ \pm 0.0029$  & $ \pm 0.0026$  & $ \pm 0.0038$  & $ \pm 0.0043$  & $ \pm 0.05$  & \\ 
 & \textbf{no} $\bm\aml$ \textbf{prior} & $1.192$  & $3.751$  & $0.129$  & $0.086$  & $8.321$  & $0.219$  & $0.0371$  & $0.0321$  & $0.0411$  & $0.6687$  & $2.16$  & NSE*  \\ 
 & & $ \pm 0.013$  & $ \pm 0.005$  & $ \pm 0.018$  & $ \pm 0.002$  & $ \pm 0.290$  & $ \pm 0.005$  & $ \pm 0.0025$  & $ \pm 0.0021$  & $ \pm 0.0027$  & $ \pm 0.0016$  & $ \pm 0.05$  & \\ 
\hline
\end{tabular}
}
\label{tab:all1}
\end{table*}

\setcounter{table}{0}

\begin{table*}
   \centering
   \caption{... continued.}
   \resizebox{0.95\textwidth}{!} {
     \begin{tabular}{l l c c c c c c c c c c c l} 
   \hline
   \hline
     Star & $\alpha_{\rm}$ prior & $M/M_{\odot}$ & $\log T_{\rm eff}$ & $\log L/L_{\odot}$ & $\log R/R_{\odot}$ & Age & $\YO$ & $\ZO$ & $\Zs$ & $\ZoX$ & $R_{\rm BCZ}$ & $\alpha_{\rm ml}$ & sys \\
     \hline
6933899 & $\alpha_{\rm ml}=1.8$  & $1.164$  & $3.756$  & $0.393$  & $0.208$  & $7.808$  & $0.248$  & $0.0245$  & $0.0212$  & $0.0275$  & $0.6779$  & $1.80$  & ASE  \\ 
& & $ \pm 0.047$  & $ \pm 0.004$  & $ \pm 0.022$  & $ \pm 0.006$  & $ \pm 0.571$  & $ \pm 0.018$  & $ \pm 0.0021$  & $ \pm 0.0020$  & $ \pm 0.0028$  & $ \pm 0.0154$  & & \\ 
& \textbf{Gaussian} & $1.140$  & $3.760$  & $0.401$  & $0.205$  & $7.806$  & $0.259$  & $0.0237$  & $0.0207$  & $0.0273$  & $0.6650$  & $1.90$  & ASE  \\ 
& & $ \pm 0.064$  & $ \pm 0.004$  & $ \pm 0.027$  & $ \pm 0.008$  & $ \pm 0.620$  & $ \pm 0.024$  & $ \pm 0.0026$  & $ \pm 0.0024$  & $ \pm 0.0033$  & $ \pm 0.0189$  & $ \pm 0.05$  & \\ 
& \textbf{no} $\bm\aml$ \textbf{prior} & $1.131$  & $3.766$  & $0.426$  & $0.204$  & $7.553$  & $0.265$  & $0.0223$  & $0.0196$  & $0.0260$  & $0.6589$  & $2.04$  & ASE*  \\ 
& & $ \pm 0.053$  & $ \pm 0.005$  & $ \pm 0.028$  & $ \pm 0.007$  & $ \pm 0.574$  & $ \pm 0.022$  & $ \pm 0.0027$  & $ \pm 0.0025$  & $ \pm 0.0035$  & $ \pm 0.0146$  & $ \pm 0.09$  & \\ 
\hline
7680114 &  $\alpha_{\rm ml}=1.8$  & $1.156$  & $3.761$  & $0.309$  & $0.157$  & $6.084$  & $0.286$  & $0.0294$  & $0.0259$  & $0.0355$  & $0.7012$  & $1.80$  & SSE  \\ 
& & $ \pm 0.025$  & $ \pm 0.003$  & $ \pm 0.010$  & $ \pm 0.003$  & $ \pm 0.544$  & $ \pm 0.009$  & $ \pm 0.0018$  & $ \pm 0.0017$  & $ \pm 0.0025$  & $ \pm 0.0068$  & & \\ 
& \textbf{Gaussian} & $1.172$  & $3.766$  & $0.333$  & $0.159$  & $5.780$  & $0.284$  & $0.0289$  & $0.0256$  & $0.0351$  & $0.7004$  & $1.89$  & SSE  \\ 
& & $ \pm 0.027$  & $ \pm 0.004$  & $ \pm 0.017$  & $ \pm 0.003$  & $ \pm 0.567$  & $ \pm 0.010$  & $ \pm 0.0022$  & $ \pm 0.0020$  & $ \pm 0.0030$  & $ \pm 0.0076$  & $ \pm 0.05$  & \\ 
& \textbf{no} $\bm\aml$ \textbf{prior} & $1.186$  & $3.771$  & $0.356$  & $0.160$  & $5.521$  & $0.284$  & $0.0281$  & $0.0251$  & $0.0345$  & $0.6997$  & $1.99$  & SSE \\ 
& & $ \pm 0.033$  & $ \pm 0.005$  & $ \pm 0.024$  & $ \pm 0.004$  & $ \pm 0.721$  & $ \pm 0.011$  & $ \pm 0.0025$  & $ \pm 0.0023$  & $ \pm 0.0034$  & $ \pm 0.0101$  & $ \pm 0.08$  & \\ 
\hline
8006161 &  $\alpha_{\rm ml}=1.8$  & $1.052$  & $3.721$  & $-0.207$  & $-0.022$  & $2.714$  & $0.265$  & $0.0395$  & $0.0377$  & $0.0532$  & $0.6891$  & $1.80$  & SSE \\ 
& & $ \pm 0.022$  & $ \pm 0.003$  & $ \pm 0.010$  & $ \pm 0.003$  & $ \pm 0.500$  & $ \pm 0.015$  & $ \pm 0.0015$  & $ \pm 0.0015$  & $ \pm 0.0026$  & $ \pm 0.0026$  & & \\ 
& Gaussian & $1.077$  & $3.721$  & $-0.201$  & $-0.019$  & $3.220$  & $0.246$  & $0.0398$  & $0.0378$  & $0.0519$  & $0.6847$  & $1.91$  & SSE  \\ 
& & $ \pm 0.027$  & $ \pm 0.003$  & $ \pm 0.011$  & $ \pm 0.004$  & $ \pm 0.541$  & $ \pm 0.020$  & $ \pm 0.0010$  & $ \pm 0.0009$  & $ \pm 0.0019$  & $ \pm 0.0037$  & $ \pm 0.07$  & \\ 
& \textbf{no} $\bm\aml$ \textbf{prior} & $1.114$  & $3.721$  & $-0.188$  & $-0.013$  & $3.896$  & $0.219$  & $0.0400$  & $0.0377$  & $0.0499$  & $0.6791$  & $2.10$  & SSE  \\ 
& & $ \pm 0.017$  & $ \pm 0.003$  & $ \pm 0.011$  & $ \pm 0.002$  & $ \pm 0.453$  & $ \pm 0.011$  & $ \pm 0.0003$  & $ \pm 0.0003$  & $ \pm 0.0008$  & $ \pm 0.0024$  & $ \pm 0.06$  & \\ 
\hline
8228742 &  $\alpha_{\rm ml}=1.8$  & $1.214$  & $3.762$  & $0.518$  & $0.260$  & $6.584$  & $0.240$  & $0.0200$  & $0.0174$  & $0.0224$  & $0.6906$  & $1.80$  & SSE \\ 
& & $ \pm 0.021$  & $ \pm 0.002$  & $ \pm 0.002$  & $ \pm 0.003$  & $ \pm 0.200$  & $ \pm 0.014$  & $ \pm 0.0003$  & $ \pm 0.0003$  & $ \pm 0.0005$  & $ \pm 0.0025$  & & \\ 
& Gaussian & $1.248$  & $3.771$  & $0.565$  & $0.264$  & $5.868$  & $0.241$  & $0.0199$  & $0.0175$  & $0.0225$  & $0.6996$  & $1.95$  & SSE  \\ 
& & $ \pm 0.025$  & $ \pm 0.004$  & $ \pm 0.017$  & $ \pm 0.003$  & $ \pm 0.259$  & $ \pm 0.012$  & $ \pm 0.0006$  & $ \pm 0.0006$  & $ \pm 0.0008$  & $ \pm 0.0038$  & $ \pm 0.05$  & \\ 
& \textbf{no} $\bm\aml$ \textbf{prior} & $1.274$  & $3.778$  & $0.596$  & $0.266$  & $5.479$  & $0.240$  & $0.0199$  & $0.0175$  & $0.0225$  & $0.7047$  & $2.05$  & SSE  \\ 
& & $ \pm 0.027$  & $ \pm 0.004$  & $ \pm 0.017$  & $ \pm 0.003$  & $ \pm 0.261$  & $ \pm 0.014$  & $ \pm 0.0006$  & $ \pm 0.0006$  & $ \pm 0.0008$  & $ \pm 0.0039$  & $ \pm 0.06$  & \\ 
\hline
8379927 &  $\alpha_{\rm ml}=1.8$  & $1.253$  & $3.774$  & $0.184$  & $0.068$  & $1.513$  & $0.226$  & $0.0250$  & $0.0237$  & $0.0310$  & $0.7638$  & $1.80$  & NSE* \\ 
& & $ \pm 0.011$  & $ \pm 0.001$  & $ \pm 0.007$  & $ \pm 0.001$  & $ \pm 0.231$  & $ \pm 0.003$  & $ \pm 0.0003$  & $ \pm 0.0003$  & $ \pm 0.0005$  & $ \pm 0.0039$  & & \\ 
& Gaussian & $1.258$  & $3.778$  & $0.201$  & $0.068$  & $1.511$  & $0.227$  & $0.0246$  & $0.0233$  & $0.0305$  & $0.7651$  & $1.86$  & NSE*  \\ 
& & $ \pm 0.016$  & $ \pm 0.004$  & $ \pm 0.018$  & $ \pm 0.002$  & $ \pm 0.248$  & $ \pm 0.004$  & $ \pm 0.0014$  & $ \pm 0.0014$  & $ \pm 0.0019$  & $ \pm 0.0044$  & $ \pm 0.06$  & \\ 
& \textbf{no} $\bm\aml$ \textbf{prior} & $1.262$  & $3.797$  & $0.279$  & $0.069$  & $1.624$  & $0.231$  & $0.0204$  & $0.0191$  & $0.0249$  & $0.7750$  & $2.18$  & NSE*  \\ 
& & $ \pm 0.017$  & $ \pm 0.007$  & $ \pm 0.029$  & $ \pm 0.002$  & $ \pm 0.231$  & $ \pm 0.004$  & $ \pm 0.0014$  & $ \pm 0.0014$  & $ \pm 0.0019$  & $ \pm 0.0060$  & $ \pm 0.13$  & \\ 
\hline
8760414 &  $\bm{\aml=1.8}$  & $0.839$  & $3.775$  & $0.084$  & $0.016$  & $11.400$  & $0.245$  & $0.0050$  & $0.0038$  & $0.0046$  & $0.7212$  & $1.80$  & SSE  \\ 
& & $ \pm 0.013$  & $ \pm 0.002$  & $ \pm 0.014$  & $ \pm 0.002$  & $ \pm 0.873$  & $ \pm 0.002$  & $ \pm < 0.0001$  & $ \pm < 0.0001$  & $ \pm < 0.0001$  & $ \pm 0.0090$  & & \\ 
& \textbf{Gaussian} & $0.838$  & $3.775$  & $0.084$  & $0.016$  & $11.426$  & $0.245$  & $0.0050$  & $0.0038$  & $0.0046$  & $0.7209$  & $1.80$  & SSE  \\ 
& & $ \pm 0.013$  & $ \pm 0.002$  & $ \pm 0.014$  & $ \pm 0.002$  & $ \pm 0.886$  & $ \pm 0.002$  & $ \pm < 0.0001$  & $ \pm < 0.0001$  & $ \pm < 0.0001$  & $ \pm 0.0092$  & $ \pm 0.02$  & \\ 
& \textbf{no} $\bm\aml$ \textbf{prior} & $0.862$  & $3.789$  & $0.147$  & $0.020$  & $10.511$  & $0.245$  & $0.0050$  & $0.0039$  & $0.0048$  & $0.7181$  & $2.25$  & SSE  \\ 
& & $ \pm 0.015$  & $ \pm 0.008$  & $ \pm 0.037$  & $ \pm 0.003$  & $ \pm 0.706$  & $ \pm 0.001$  & $ \pm < 0.0001$  & $ \pm < 0.0001$  & $ \pm 0.0001$  & $ \pm 0.0049$  & $ \pm 0.26$  & \\ 
\hline
10516096 &  $\alpha_{\rm ml}=1.8$  & $1.185$  & $3.765$  & $0.338$  & $0.163$  & $6.049$  & $0.258$  & $0.0244$  & $0.0213$  & $0.0280$  & $0.7128$  & $1.80$  & NSE  \\ 
& & $ \pm 0.017$  & $ \pm 0.002$  & $ \pm 0.009$  & $ \pm 0.002$  & $ \pm 0.461$  & $ \pm 0.016$  & $ \pm 0.0016$  & $ \pm 0.0016$  & $ \pm 0.0024$  & $ \pm 0.0060$  & & \\ 
& Gaussian & $1.210$  & $3.772$  & $0.374$  & $0.166$  & $5.854$  & $0.247$  & $0.0229$  & $0.0201$  & $0.0261$  & $0.7160$  & $1.92$  & NSE  \\ 
& & $ \pm 0.021$  & $ \pm 0.004$  & $ \pm 0.020$  & $ \pm 0.003$  & $ \pm 0.533$  & $ \pm 0.019$  & $ \pm 0.0025$  & $ \pm 0.0023$  & $ \pm 0.0034$  & $ \pm 0.0079$  & $ \pm 0.06$  & \\ 
& \textbf{no} $\bm\aml$ \textbf{prior} & $1.240$  & $3.781$  & $0.419$  & $0.170$  & $5.500$  & $0.238$  & $0.0213$  & $0.0189$  & $0.0243$  & $0.7188$  & $2.11$  & NSE  \\ 
& & $ \pm 0.018$  & $ \pm 0.005$  & $ \pm 0.022$  & $ \pm 0.002$  & $ \pm 0.500$  & $ \pm 0.013$  & $ \pm 0.0022$  & $ \pm 0.0020$  & $ \pm 0.0028$  & $ \pm 0.0072$  & $ \pm 0.09$  & \\ 
\hline
10963065 &  $\alpha_{\rm ml}=1.8$  & $1.122$  & $3.778$  & $0.259$  & $0.097$  & $5.035$  & $0.252$  & $0.0174$  & $0.0147$  & $0.0189$  & $0.7511$  & $1.80$  & SSE  \\ 
& & $ \pm 0.037$  & $ \pm 0.005$  & $ \pm 0.025$  & $ \pm 0.005$  & $ \pm 0.945$  & $ \pm 0.018$  & $ \pm 0.0025$  & $ \pm 0.0023$  & $ \pm 0.0032$  & $ \pm 0.0167$  & & \\ 
& \textbf{Gaussian} & $1.094$  & $3.777$  & $0.248$  & $0.094$  & $6.139$  & $0.252$  & $0.0176$  & $0.0148$  & $0.0191$  & $0.7254$  & $1.92$  & SSE  \\ 
& & $ \pm 0.038$  & $ \pm 0.005$  & $ \pm 0.026$  & $ \pm 0.005$  & $ \pm 1.090$  & $ \pm 0.18$  & $ \pm 0.0025$  & $ \pm 0.0022$  & $ \pm 0.0032$  & $ \pm 0.0198$  & $ \pm 0.07$  & \\ 
& \textbf{no} $\bm\aml$ \textbf{prior} & $1.089$  & $3.785$  & $0.278$  & $0.093$  & $6.538$  & $0.245$  & $0.0154$  & $0.0129$  & $0.0165$  & $0.7162$  & $2.15$  & SSE  \\ 
& & $ \pm 0.029$  & $ \pm 0.005$  & $ \pm 0.026$  & $ \pm 0.004$  & $ \pm 0.846$  & $ \pm 0.012$  & $ \pm 0.0013$  & $ \pm 0.0012$  & $ \pm 0.0017$  & $ \pm 0.0135$  & $ \pm 0.09$  & \\ 
\hline
11244118 &  $\alpha_{\rm ml}=1.8$  & $1.233$  & $3.751$  & $0.392$  & $0.218$  & $7.100$  & $0.265$  & $0.0388$  & $0.0345$  & $0.0470$  & $0.6830$  & $1.80$  & SSE  \\ 
& & $ \pm 0.053$  & $ \pm 0.007$  & $ \pm 0.038$  & $ \pm 0.006$  & $ \pm 1.232$  & $ \pm 0.014$  & $ \pm 0.0026$  & $ \pm 0.0025$  & $ \pm 0.0036$  & $ \pm 0.0267$  & & \\ 
& Gaussian & $1.299$  & $3.752$  & $0.412$  & $0.227$  & $7.633$  & $0.231$  & $0.0400$  & $0.0359$  & $0.0471$  & $0.6557$  & $2.01$  & NSE*  \\ 
& & $ \pm 0.004$  & $ \pm 0.002$  & $ \pm 0.008$  & $ \pm 0.000$  & $ \pm 0.222$  & $ \pm 0.005$  & $ \pm 0.0001$  & $ \pm 0.0001$  & $ \pm 0.0004$  & $ \pm 0.0017$  & $ \pm 0.03$  & \\ 
& \textbf{no} $\bm\aml$ \textbf{prior} & $1.291$  & $3.759$  & $0.438$  & $0.226$  & $6.962$  & $0.247$  & $0.0400$  & $0.0360$  & $0.0483$  & $0.6603$  & $2.09$  & NSE*  \\ 
& & $ \pm 0.004$  & $ \pm 0.003$  & $ \pm 0.011$  & $ \pm 0.000$  & $ \pm 0.292$  & $ \pm 0.007$  & $ \pm < 0.0001$  & $ \pm 0.0001$  & $ \pm 0.0005$  & $ \pm 0.0021$  & $ \pm 0.04$  & \\ 
\hline
11713510 &  $\bm{\aml=1.8}$  & $1.025$  & $3.772$  & $0.441$  & $0.200$  & $7.135$  & $0.291$  & $0.0139$  & $0.0115$  & $0.0153$  & $0.6992$  & $1.80$  & SSE  \\ 
& & $ \pm 0.019$  & $ \pm 0.003$  & $ \pm 0.013$  & $ \pm 0.003$  & $ \pm 0.508$  & $ \pm 0.019$  & $ \pm 0.0022$  & $ \pm 0.0019$  & $ \pm 0.0028$  & $ \pm 0.0121$  & & \\ 
& \textbf{Gaussian} & $1.031$  & $3.773$  & $0.447$  & $0.201$  & $7.042$  & $0.294$  & $0.0145$  & $0.0121$  & $0.0162$  & $0.6930$  & $1.85$  & SSE  \\ 
& & $ \pm 0.022$  & $ \pm 0.003$  & $ \pm 0.013$  & $ \pm 0.003$  & $ \pm 0.424$  & $ \pm 0.017$  & $ \pm 0.0019$  & $ \pm 0.0017$  & $ \pm 0.0025$  & $ \pm 0.0121$  & $ \pm 0.05$  & \\ 
& \textbf{no} $\bm\aml$ \textbf{prior} & $1.082$  & $3.775$  & $0.467$  & $0.208$  & $6.705$  & $0.290$  & $0.0182$  & $0.0157$  & $0.0213$  & $0.6829$  & $1.99$  & SSE*  \\ 
& & $ \pm 0.071$  & $ \pm 0.003$  & $ \pm 0.025$  & $ \pm 0.010$  & $ \pm 0.522$  & $ \pm 0.016$  & $ \pm 0.0057$  & $ \pm 0.0055$  & $ \pm 0.0077$  & $ \pm 0.0138$  & $ \pm 0.15$  & \\ 
\hline
12009504 &  $\alpha_{\rm ml}=1.8$  & $1.238$  & $3.773$  & $0.360$  & $0.157$  & $4.558$  & $0.252$  & $0.0239$  & $0.0207$  & $0.0270$  & $0.7451$  & $1.80$  & SSE  \\ 
& & $ \pm 0.034$  & $ \pm 0.003$  & $ \pm 0.017$  & $ \pm 0.004$  & $ \pm 0.488$  & $ \pm 0.019$  & $ \pm 0.0021$  & $ \pm 0.0019$  & $ \pm 0.0028$  & $ \pm 0.0122$  & & \\ 
& Gaussian & $1.245$  & $3.779$  & $0.386$  & $0.158$  & $4.487$  & $0.250$  & $0.0223$  & $0.0195$  & $0.0253$  & $0.7419$  & $1.93$  & SSE  \\ 
& & $ \pm 0.028$  & $ \pm 0.004$  & $ \pm 0.018$  & $ \pm 0.003$  & $ \pm 0.480$  & $ \pm 0.018$  & $ \pm 0.0025$  & $ \pm 0.0022$  & $ \pm 0.0033$  & $ \pm 0.0112$  & $ \pm 0.05$  & \\ 
& \textbf{no} $\bm\aml$ \textbf{prior} & $1.253$  & $3.786$  & $0.416$  & $0.159$  & $4.332$  & $0.249$  & $0.0205$  & $0.0180$  & $0.0234$  & $0.7410$  & $2.07$  & SSE  \\ 
& & $ \pm 0.026$  & $ \pm 0.005$  & $ \pm 0.021$  & $ \pm 0.003$  & $ \pm 0.367$  & $ \pm 0.016$  & $ \pm 0.0016$  & $ \pm 0.0014$  & $ \pm 0.0021$  & $ \pm 0.0103$  & $ \pm 0.08$  & \\ 
\hline
12258514 &  $\alpha_{\rm ml}=1.8$  & $1.250$  & $3.769$  & $0.440$  & $0.206$  & $5.564$  & $0.250$  & $0.0256$  & $0.0224$  & $0.0294$  & $0.7291$  & $1.80$  & SSE  \\ 
& & $ \pm 0.039$  & $ \pm 0.004$  & $ \pm 0.020$  & $ \pm 0.004$  & $ \pm 0.939$  & $ \pm 0.018$  & $ \pm 0.0021$  & $ \pm 0.0019$  & $ \pm 0.0028$  & $ \pm 0.0137$  & & \\ 
& \textbf{Gaussian} & $1.227$  & $3.769$  & $0.436$  & $0.204$  & $6.342$  & $0.246$  & $0.0255$  & $0.0225$  & $0.0293$  & $0.7086$  & $1.90$  & SSE  \\ 
& & $ \pm 0.038$  & $ \pm 0.003$  & $ \pm 0.017$  & $ \pm 0.004$  & $ \pm 0.823$  & $ \pm 0.019$  & $ \pm 0.0024$  & $ \pm 0.0023$  & $ \pm 0.0035$  & $ \pm 0.0138$  & $ \pm 0.04$  & \\ 
& \textbf{no} $\bm\aml$ \textbf{prior} & $1.217$  & $3.771$  & $0.445$  & $0.203$  & $6.445$  & $0.245$  & $0.0242$  & $0.0213$  & $0.0278$  & $0.7046$  & $1.96$  & SSE  \\ 
& & $ \pm 0.041$  & $ \pm 0.004$  & $ \pm 0.020$  & $ \pm 0.005$  & $ \pm 0.701$  & $ \pm 0.018$  & $ \pm 0.0029$  & $ \pm 0.0027$  & $ \pm 0.0039$  & $ \pm 0.0134$  & $ \pm 0.06$  & \\ 
\hline
\end{tabular}
}
\label{tab:all2}
\end{table*}

\setcounter{table}{0}

\begin{table*}
   \centering
   \caption{... continued.}
   \resizebox{0.95\textwidth}{!} {
     \begin{tabular}{l l c c c c c c c c c c c l} 
   \hline
   \hline
     Star & $\alpha_{\rm}$ prior & $M/M_{\odot}$ & $\log T_{\rm eff}$ & $\log L/L_{\odot}$ & $\log R/R_{\odot}$ & Age & $\YO$ & $\ZO$ & $\Zs$ & $\ZoX$ & $R_{\rm BCZ}$ & $\alpha_{\rm ml}$ & sys \\
 \hline
16CygA &  $\alpha_{\rm ml}=1.8$  & $1.054$  & $3.762$  & $0.173$  & $0.086$  & $6.441$  & $0.291$  & $0.0250$  & $0.0214$  & $0.0291$  & $0.7027$  & $1.80$  & SSE  \\ 
& & $ \pm 0.010$  & $ \pm 0.001$  & $ \pm 0.006$  & $ \pm 0.001$  & $ \pm 0.363$  & $ \pm 0.006$  & $ \pm < 0.0001$  & $ \pm 0.0002$  & $ \pm 0.0004$  & $ \pm 0.0036$  & & \\ 
& Gaussian & $1.095$  & $3.765$  & $0.196$  & $0.092$  & $7.055$  & $0.280$  & $0.0281$  & $0.0247$  & $0.0337$  & $0.6730$  & $2.13$  & SSE  \\ 
& & $ \pm 0.016$  & $ \pm 0.005$  & $ \pm 0.023$  & $ \pm 0.002$  & $ \pm 0.375$  & $ \pm 0.012$  & $ \pm 0.0024$  & $ \pm 0.0021$  & $ \pm 0.0033$  & $ \pm 0.0056$  & $ \pm 0.06$  & \\ 
& \textbf{no} $\bm\aml$ \textbf{prior} & $1.114$  & $3.771$  & $0.225$  & $0.095$  & $6.647$  & $0.269$  & $0.0250$  & $0.0220$  & $0.0295$  & $0.6795$  & $2.20$  & SSE  \\ 
& & $ \pm 0.009$  & $ \pm 0.001$  & $ \pm 0.004$  & $ \pm 0.001$  & $ \pm 0.206$  & $ \pm 0.006$  & $ \pm 0.0003$  & $ \pm 0.0002$  & $ \pm 0.0005$  & $ \pm 0.0015$  & $ \pm 0.01$  & \\ 
\hline
16CygB &  $\alpha_{\rm ml}=1.8$  & $1.007$  & $3.758$  & $0.070$  & $0.043$  & $6.464$  & $0.294$  & $0.0247$  & $0.0214$  & $0.0294$  & $0.6986$  & $1.80$  & SSE  \\ 
& & $ \pm 0.006$  & $ \pm 0.002$  & $ \pm 0.007$  & $ \pm 0.001$  & $ \pm 0.250$  & $ \pm 0.004$  & $ \pm 0.0012$  & $ \pm 0.0010$  & $ \pm 0.0015$  & $ \pm 0.0035$  & & \\ 
& Gaussian & $1.023$  & $3.762$  & $0.091$  & $0.045$  & $6.532$  & $0.289$  & $0.0250$  & $0.0217$  & $0.0296$  & $0.6942$  & $1.92$  & SSE  \\ 
& & $ \pm 0.013$  & $ \pm 0.002$  & $ \pm 0.010$  & $ \pm 0.002$  & $ \pm 0.281$  & $ \pm 0.007$  & $ \pm 0.0001$  & $ \pm 0.0001$  & $ \pm 0.0003$  & $ \pm 0.0034$  & $ \pm 0.04$  & \\ 
& \textbf{no} $\bm\aml$ \textbf{prior} & $1.076$  & $3.764$  & $0.116$  & $0.054$  & $9.279$  & $0.234$  & $0.0250$  & $0.0214$  & $0.0274$  & $0.6621$  & $2.40$  & SSE  \\ 
& & $ \pm 0.012$  & $ \pm 0.002$  & $ \pm 0.009$  & $ \pm 0.002$  & $ \pm 0.473$  & $ \pm 0.005$  & $ \pm 0.0001$  & $ \pm 0.0001$  & $ \pm 0.0003$  & $ \pm 0.0035$  & $ \pm 0.00$  & \\ 
\hline
Kepler36 &  $\bm{\aml=1.8}$  & $1.113$  & $3.771$  & $0.475$  & $0.220$  & $6.923$  & $0.256$  & $0.0150$  & $0.0124$  & $0.0159$  & $0.7059$  & $1.80$  & NSE  \\ 
& & $ \pm 0.035$  & $ \pm 0.003$  & $ \pm 0.015$  & $ \pm 0.005$  & $ \pm 0.372$  & $ \pm 0.018$  & $ \pm 0.0004$  & $ \pm 0.0004$  & $ \pm 0.0006$  & $ \pm 0.0121$  & & \\ 
& \textbf{Gaussian} & $1.118$  & $3.771$  & $0.480$  & $0.221$  & $6.870$  & $0.255$  & $0.0150$  & $0.0125$  & $0.0159$  & $0.7058$  & $1.82$  & NSE  \\ 
& & $ \pm 0.035$  & $ \pm 0.003$  & $ \pm 0.017$  & $ \pm 0.005$  & $ \pm 0.386$  & $ \pm 0.018$  & $ \pm 0.0004$  & $ \pm 0.0004$  & $ \pm 0.0006$  & $ \pm 0.0122$  & $ \pm 0.04$  & \\ 
& \textbf{no} $\bm\aml$ \textbf{prior} & $1.123$  & $3.773$  & $0.486$  & $0.222$  & $6.792$  & $0.254$  & $0.0150$  & $0.0125$  & $0.0160$  & $0.7058$  & $1.85$  & NSE  \\ 
& & $ \pm 0.036$  & $ \pm 0.004$  & $ \pm 0.021$  & $ \pm 0.005$  & $ \pm 0.409$  & $ \pm 0.018$  & $ \pm 0.0004$  & $ \pm 0.0004$  & $ \pm 0.0006$  & $ \pm 0.0122$  & $ \pm 0.06$  & \\ 
\hline
\end{tabular}
}
\label{tab:all3}
\end{table*}

\end{document}